\documentclass[11pt]{article}

\newsavebox{\foobox}
\newcommand{\slantbox}[2][0]{\mbox{%
        \sbox{\foobox}{#2}%
        \hskip\wd\foobox
        \pdfsave
        \pdfsetmatrix{1 0 #1 1}%
        \llap{\usebox{\foobox}}%
        \pdfrestore
}}
\newcommand\unslant[2][-.25]{\slantbox[#1]{$#2$}}

\newcommand{\mdelta}{\text{\unslant[-.18]\delta}}

\usepackage[left=2cm, right=2cm, top=2.5cm, bottom=2.5cm]{geometry}
\usepackage[x11names]{xcolor}
\usepackage{fancyhdr, amssymb, cancel, amsmath, graphicx, pgfplots, tikz}
\usepackage{isomath}

\usetikzlibrary{shadows}

\newcommand{\stylecolor}{green!40!black}

\usepackage[labelfont={bf,sf, color=\stylecolor}, margin={1.5cm,0cm}]{caption}

\usepackage[colorlinks=true, urlcolor=\stylecolor!70!white, linkcolor=\stylecolor, citecolor=\stylecolor!70!white, hyperindex=true, linktocpage=true]{hyperref}

\usepackage{amsthm}
\newtheoremstyle{theor}{10pt}{10pt}{\addtolength{\leftskip}{2.5em}}{}{\sffamily \bfseries \color{green!50!black}}{:}{.5em}{}
\theoremstyle{theor}

\usepackage[explicit]{titlesec}

\newcommand*\sectionlabel{}
\titleformat{\section}
  {\gdef\sectionlabel{}
   \Large\bfseries\scshape}
  {\gdef\sectionlabel{\thesection }}{0pt}
  {\begin{tikzpicture}[remember picture,overlay]
       \end{tikzpicture}
  }
\titlespacing*{\section}{0pt}{0pt}{0pt}

\newcommand*\subsectionlabel{}
\titleformat{\subsection}
  {\gdef\subsectionlabel{}
   \large\bfseries\scshape}
  {\gdef\subsectionlabel{\thesubsection  }}{0pt}
  {\begin{tikzpicture}[remember picture,overlay]
    	\draw (-0.15, 0.02) node[right] {\color{\stylecolor} \textsf{\subsectionlabel}};
	\draw (1.25, 0) node[right] {\color{\stylecolor} \textsf{#1}};
	\fill[color=\stylecolor] (1,-0.25) rectangle (1.1, 0.25);
       \end{tikzpicture}
  }
\titlespacing*{\subsection}{0pt}{10pt}{10pt}

\newcommand*\subsubsectionlabel{}
\titleformat{\subsubsection}
  {\gdef\subsubsectionlabel{}
   \bfseries\scshape}
  {\gdef\subsubsectionlabel{\thesubsubsection.\ \  }}{0pt}
  {\begin{tikzpicture}[remember picture,overlay]
    	\draw (-0.15, 0) node[right] {\color{\stylecolor} \textsf{\subsubsectionlabel#1}};
       \end{tikzpicture}
  }
\titlespacing*{\subsubsection}{0pt}{7pt}{7pt}

\pgfplotsset{every axis legend/.append style={at={(1.02,1)},anchor=north west}}

\begin{document}

\allowdisplaybreaks

\pagestyle{fancy}
\renewcommand{\headrulewidth}{0pt}
\fancyhead{}

\fancyfoot{}
\fancyfoot[C] {\textsf{\textbf{\thepage}}}

\begin{equation*}
\begin{tikzpicture}
\draw (\textwidth, 0) node[text width = \textwidth, right] {\color{white} easter egg};
\end{tikzpicture}
\end{equation*}

\begin{equation*}
\begin{tikzpicture}
\draw (0.5\textwidth, -3) node[text width = \textwidth] {\huge  \textsf{\textbf{Hard combinatorial problems and minor embeddings\\ \vspace{0.07in}   on lattice graphs}} };
\end{tikzpicture}
\end{equation*}
\begin{equation*}
\begin{tikzpicture}
\draw (0.5\textwidth, 0.1) node[text width=\textwidth] {\large \color{black} \textsf{Andrew Lucas}};
\draw (0.5\textwidth, -0.5) node[text width=\textwidth] {\small \textsf{Department of Physics, Stanford University, Stanford, CA 94305, USA}};
\draw (0.5\textwidth, -1.0) node[text width=\textwidth] {\small \textsf{D-Wave Systems Inc., Burnaby, BC, Canada}};
\end{tikzpicture}
\end{equation*}
\begin{equation*}
\begin{tikzpicture}
\draw (0, -13.1) node[right, text width=0.5\paperwidth] {\texttt{ajlucas@stanford.edu}};
\draw (\textwidth, -13.1) node[left] {\textsf{\today}};
\end{tikzpicture}
\end{equation*}
\begin{equation*}
\begin{tikzpicture}
\draw[very thick, color=\stylecolor] (0.0\textwidth, -5.75) -- (0.99\textwidth, -5.75);
\draw (0.12\textwidth, -6.25) node[left] {\color{\stylecolor}  \textsf{\textbf{Abstract:}}};
\draw (0.53\textwidth, -6) node[below, text width=0.8\textwidth, text justified] {\small Today, hardware constraints are an important limitation on quantum adiabatic optimization algorithms.   Firstly, computational problems must be formulated as quadratic unconstrained binary optimization (QUBO) in the presence of noisy coupling constants.   Secondly, the interaction graph of the QUBO must have an effective minor embedding into a two-dimensional nonplanar lattice graph.   We describe new strategies for constructing \linebreak QUBOs for NP-complete/hard combinatorial problems that address both of these challenges.   Our results include asymptotically improved embeddings for number partitioning, filling knapsacks, graph coloring, and finding Hamiltonian cycles.   These embeddings can be also be found with reduced computational effort.   Our new embedding for number partitioning may be more effective on next-generation  hardware.};
\end{tikzpicture}
\end{equation*}

\tableofcontents

\titleformat{\section}
  {\gdef\sectionlabel{}
   \Large\bfseries\scshape}
  {\gdef\sectionlabel{\thesection }}{0pt}
  {\begin{tikzpicture}[remember picture,overlay]
	\draw (1, 0) node[right] {\color{\stylecolor} \textsf{#1}};
	\fill[color=\stylecolor] (0,-0.37) rectangle (0.7, 0.37);
	\draw (0.35, 0) node {\color{white} \textsf{\sectionlabel}};
       \end{tikzpicture}
  }
\titlespacing*{\section}{0pt}{15pt}{15pt}

\begin{equation*}
\begin{tikzpicture}
\draw[very thick, color=\stylecolor] (0.0\textwidth, -5.75) -- (0.99\textwidth, -5.75);
\end{tikzpicture}
\end{equation*}

\section{Introduction}
Since 1972, it has been known that canonical combinatorial problems with numerous industrial applications --- including but not limited to graph coloring, job scheduling, number partitioning and the traveling salesman problem --- are NP-complete: it is widely believed that they cannot effectively be solved on classical computers \cite{karp}.  Exponential runtime is required to solve many kinds of  NP-complete problems with existing classical methods: any improved algorithms are of great interest.  Ever since the proposal that a quantum annealer may outperform classical computers on certain NP-complete problems \cite{nishimori, farhi}, there has been an enormous effort to realize such quantum computational speed-up, culminating in the development of increasingly larger commercial quantum annealing (QA) devices \cite{dwave}, which appear to be thermally-assisted quantum annealers \cite{boixo14, pudenz, lanting}.

The standard QA device employs a technique called quantum adiabatic optimization \cite{bapst}, based on the adiabatic principle of quantum mechanics.  Let $H(t)$ be a time-dependent Hamiltonian, and let $|\Psi(t)\rangle$ denote the state of the quantum system, which evolves according to $\frac{\mathrm{d}}{\mathrm{d}t}|\Psi\rangle = -\mathrm{i}H|\Psi\rangle$.    Then if $|\Psi(0)\rangle$ is in the ground state of $H(0)$ (it is an eigenvector of minimal eigenvalue),  $|\Psi(t)\rangle$ remains in the ground state of $H(t)$ so long as $\mathrm{d}H/\mathrm{d}t$ is sufficiently small.    Using this adiabatic principle, we may solve combinatorial problems as follows.   Let $H_{\mathrm{D}}$ denote a Hamiltonian whose ground state is easy to prepare, and let $H_{\mathrm{P}}$ denote a Hamiltonian whose ground states are in one-to-one correspondence with the solutions to a combinatorial problem.   Then, prepare a quantum system such that $|\Psi(0)\rangle$ is in the ground state of $H_{\mathrm{D}}$, and evolve the system with a time-dependent Hamiltonian \begin{equation}
H(t) = \left(1-\frac{t}{\tau}\right)H_{\mathrm{D}} + \frac{t}{\tau} H_{\mathrm{P}}
\end{equation}
for time $\tau$.     If $\tau$ is sufficiently large, then $|\Psi(\tau)\rangle$ is highly likely to be in the ground state of $H_{\mathrm{P}}$, therefore encoding the solution to our problem.    

There are two significant challenges facing QA as a new algorithm for solving NP-complete problems.   Firstly, and most importantly, does QA actually provide any advantage over a classical algorithm?   Largely in highly tuned problems \cite{denchev, mandra16, mcgeoch}, there is some evidence for quantum speedup \cite{ronnow, heim, hen15, munozbauza}.   But there are also theoretical arguments that \cite{bapst, altshuler, dickson, hen11, farhi12} 
\begin{equation}
\tau \sim \exp\left[\alpha N^\beta\right]
\end{equation}
where $\alpha$ and $\beta$ are O(1) constants, and $N$ is the ``size" of the combinatorial problem.   This exponential scaling has been observed in experiments \cite{ronnow, heim, hen15, munozbauza}, and a simple cartoon of why $\beta=1$ is sensible is given in Appendix \ref{app:gap}.   If this exponential scaling is true, then it is imperative to minimize both $\alpha$ and $\beta$.

Another important problem -- and the one we will focus on in this paper -- is how to actually \emph{construct} $H_{\mathrm{P}}$.   Existing hardware allows us to encode quadratic unconstrained binary optimization (QUBO) problems:  \cite{boros} \begin{equation}
H_{\mathrm{P}} = \sum_{i,j=1}^N h_{i,j} Z_i Z_j,  \label{eq:QUBO}
\end{equation}
where $Z_i$ denote Pauli matrices that measure whether spin $i$ is up or down:  $Z_i|\uparrow_i\rangle = |\uparrow_i\rangle$,  $Z_i |\downarrow_i\rangle = 0$.    So long as we may tune $N$ and $h_{i,j}$, there are (infinitely) many choices of $H_{\mathrm{P}}$ for any given problem.   Ultimately, the best choice of $H_{\mathrm{P}}$ depends on the $\tau$ required for each.   However, nobody knows how to practically compute $\tau$ from first principles.   So far, a good rule of thumb has been that minimizing the number of bits $N$ in the QUBO will decrease $\tau$ \cite{rieffel}, although this is not always the case \cite{bian16}.

Unfortunately, not all choices of $h_{i,j}$ can effectively be encoded in hardware.   Current hardware embeds 
\begin{equation}
H_{\mathrm{P}} = \sum_{I\sim J} \mathfrak{h}_{IJ} Z_I Z_J,  \label{eq:QUBOhardware}
\end{equation}
where $I\sim J$ denotes a sum over only the variables which are neighbors on the Chimera graph \cite{bunyk}, which is a  (finite subset of a) nonplanar two-dimensional lattice.   The Chimera graph is employed in the hardware used by D-Wave Systems, and will be described below.   As this is a lattice, the degree of the vertices does \emph{not} scale with the number $N^\prime$ of bits $Z_I$.   Only $\mathrm{O}(N^\prime )$ $\mathfrak{h}_{IJ}$ couplings are allowed, in contrast to the $\mathrm{O}(N^2)$ $h_{i,j}$ allowed in (\ref{eq:QUBO}).    Therefore, one must convert $h_{i,j}$ into $\mathfrak{h}_{IJ}$.    The standard process for doing this is to find a minor embedding of a graph $G$, whose edge set is defined by the nonvanishing $h_{i,j}$ couplings, into the Chimera graph.   This means that multiple of the $Z_I$ in (\ref{eq:QUBOhardware}) will be used to physically represent the same \emph{logical} bit  $Z_i$ in the combinatorial problem (\ref{eq:QUBO}).    The worst case scenario is that $\mathrm{O}(N)$ physical bits $Z_I$ are required for every logical bit $Z_i$:  $N^\prime \sim N^2$.    A second hardware issue is the inevitable presence of noise, which requires that each non-vanishing $\mathfrak{h}_{IJ}$ be comparable in magnitude \cite{bunyk}.   Most existing strategies \cite{lucas} for mapping combinatorial problems into QUBOs lead to QUBOs that do not admit an efficient minor embedding into a lattice \cite{rieffel}, or that have a wide range in couplings $h_{i,j}$ that are incompatible with the inevitable presence of noise in $\mathfrak{h}_{IJ}$.    Furthermore, even if an efficient minor embedding exists, it may be exponentially hard to find \cite{choi1, choi2, adler}.

The purpose of this paper is to present new QUBOs for a number of classic combinatorial problems that address all of these issues.   Firstly, they embed more effectively into lattice graphs.  Secondly, they do not require exquisite control over coupling constants.   Finally, all of the embeddings that we discuss can be found significantly faster than previously;  in some cases, the minor embedding is explicitly given, so there is no computational time to find it.    Although we anticipate the methods developed in this paper can find broad applicability, here we will focus on a select set of classic combinatorial problems: the number partitioning problem, the knapsack problem, the graph coloring problem and the Hamiltonian cycles problem.   In each case, we find a parametrically better way to write the problem in the form (\ref{eq:QUBOhardware}) than has been known to date.   In particular, the scaling of $N^\prime $ with $N$ that we obtain  is an asymptotic improvement over existing methods and, for some problems, is provably optimal.

The rest of this paper is organized as follows.  Section \ref{sec:2d} gives some mathematical definitions of lattice graphs and minor embeddings, and reformulates the hardware constraints above more precisely.  Sections \ref{sec:tree} and \ref{sec:binary} describe new embeddings for simple computational problems that will prove crucial in our more sophisticated constructions.   Section \ref{sec:NK} describes the new embeddings for number partitioning and knapsack problems, Sections \ref{sec:combgraphs} and \ref{sec:color} describe the graph coloring problem, and Section \ref{sec:ham} describes the Hamiltonian cycles problem.  We summarize our findings in Section \ref{sec:conc}.  Appendices contain further technical details.

\section{Lattice Graphs (in Two Dimensions)}\label{sec:2d}
Our first goal is to reintroduce the minor embedding problem described in the introduction, now using more precise language.   We define an undirected graph $G = (V,E)$ as a set of vertices $V$ and a set of unoriented edges $E$ between these vertices: e.g., if vertices $i$ and $j$ are connected by an edge, $(ij) = (ji) \in E$.     Note that $G$ is undirected because the matrix of couplings $h_{i,j}$ in (\ref{eq:QUBO}) is necessarily symmetric.   We will say that a QUBO (\ref{eq:QUBO}) is defined on graph $G$ if for $i\ne j$,  $h_{i,j} \ne 0$ if and only if $(ij)\in E$.   The QUBOs for hard combinatorial problems of interest are often defined on ``infinite dimensional" graphs $G$, which look nothing like a two-dimensional lattice \cite{lucas}.  

QA on ``large" problems have only been implemented experimentally on QUBOs that can be defined on finite subgraphs of a two-dimensional lattice.  Mathematically, we  define a two dimensional lattice as an infinite undirected graph $\Lambda_0 = (V_{\Lambda_0},E_{\Lambda_0})$ whose automorphism group $\mathrm{Aut}(\Lambda_0)$, defined as \begin{equation}
\mathrm{Aut}(\Lambda_0) = \lbrace \sigma: V_{\Lambda_0} \rightarrow V_{\Lambda_0} \; | \; (\sigma(v_1)\sigma(v_2))\in E_{\Lambda_0} \iff (v_1v_2) \in E_{\Lambda_0} \rbrace,
\end{equation}
contains a $\mathbb{Z}\times \mathbb{Z}$ subgroup.   
In this paper, we will focus on lattices $\Lambda_0$ which take a particularly simple form:  \begin{equation}
\Lambda_0 = \left( \bigcup_{i,j \in \mathbb{Z}} V_{i,j}, \; \bigcup_{i,j \in \mathbb{Z}} E_{i,j} \cup E^{\mathrm{h}}_{i,j} \cup E^{\mathrm{v}}_{i,j} \right),
\end{equation}
where $V_{i,j} = \lbrace v_{i,j}^1, \ldots, v^n_{i,j}\rbrace$ is a set of $n$ vertices.   Let us define a set of three matrices $A^{ab} = A^{ba}$, $A_{\mathrm{h}}^{ab}$, and $A_{\mathrm{v}}^{ab}$;  we then define \begin{subequations}\begin{align}
E_{i,j} &= \bigcup_{a<b \; : \;  A^{ab}=1} \lbrace (v_{i,j}^av_{i,j}^b) \rbrace, \\
E^{\mathrm{h}}_{i,j} &= \bigcup_{a,b \; :\;  A_{\mathrm{h}}^{ab}=1} \lbrace (v_{i,j}^av_{i+1,j}^b) \rbrace, \\
E^{\mathrm{v}}_{i,j} &= \bigcup_{a,b \; : \;   A_{\mathrm{v}}^{ab}=1} \lbrace (v_{i,j}^av_{i,j+1}^b) \rbrace.
\end{align}\end{subequations}
The $\mathbb{Z}\times \mathbb{Z}$ subgroup of $\mathrm{Aut}(\Lambda_0)$ corresponds to $v_{i,j}^a \rightarrow v_{i+m, j+n}^a$ for $m,n\in \mathbb{Z}$.    We are interested in subgraphs of the form \begin{equation}
\Lambda = \left( \bigcup_{1\le i,j\le L} V_{i,j},  \;   \left\lbrace \bigcup_{1\le i,j\le L} E_{i,j} \right\rbrace \cup \left\lbrace \bigcup_{1\le i\le L-1, 1\le j\le L} E^{\mathrm{h}}_{i,j} \right\rbrace  \cup \left\lbrace \bigcup_{1\le i\le L, 1\le j\le L-1} E^{\mathrm{v}}_{i,j} \right\rbrace  \right)  . \label{eq:lattice}
\end{equation}
We  refer to subgraphs $(V_{i,j}, E_{i,j})$ as cells.   To avoid saying ``subgraph of $L\times L$ cells" repeatedly, we will call $\Lambda$ an $L\times L$ lattice in the rest of this paper.    Finally, we refer to $L$ as the length of the lattice, and define $e = |E_{i,j}|$, $e_{\mathrm{h}} = |E^{\mathrm{h}}_{i,j}|$ and $e_{\mathrm{v}} = |E^{\mathrm{v}}_{i,j}|$.   Note that the total number of vertices in $\Lambda$ is \begin{equation}
N^\prime = nL^2.
\end{equation}
The total number of edges also scales as $L^2$;  the average degree $\bar k$ of a vertex in $\Lambda$ is bounded by \begin{equation}
\bar k \le \frac{2}{n}\left(e + e_{\mathrm{h}}+e_{\mathrm{v}}\right).
\end{equation}
In this paper, we will focus on the case where $n$, $e$, $e_{\mathrm{h}}$ and $e_{\mathrm{v}}$ are all constants which do not scale with $N$ (or any other parametrically large parameter).

\begin{figure}[t]
\centering
\includegraphics[width=5in]{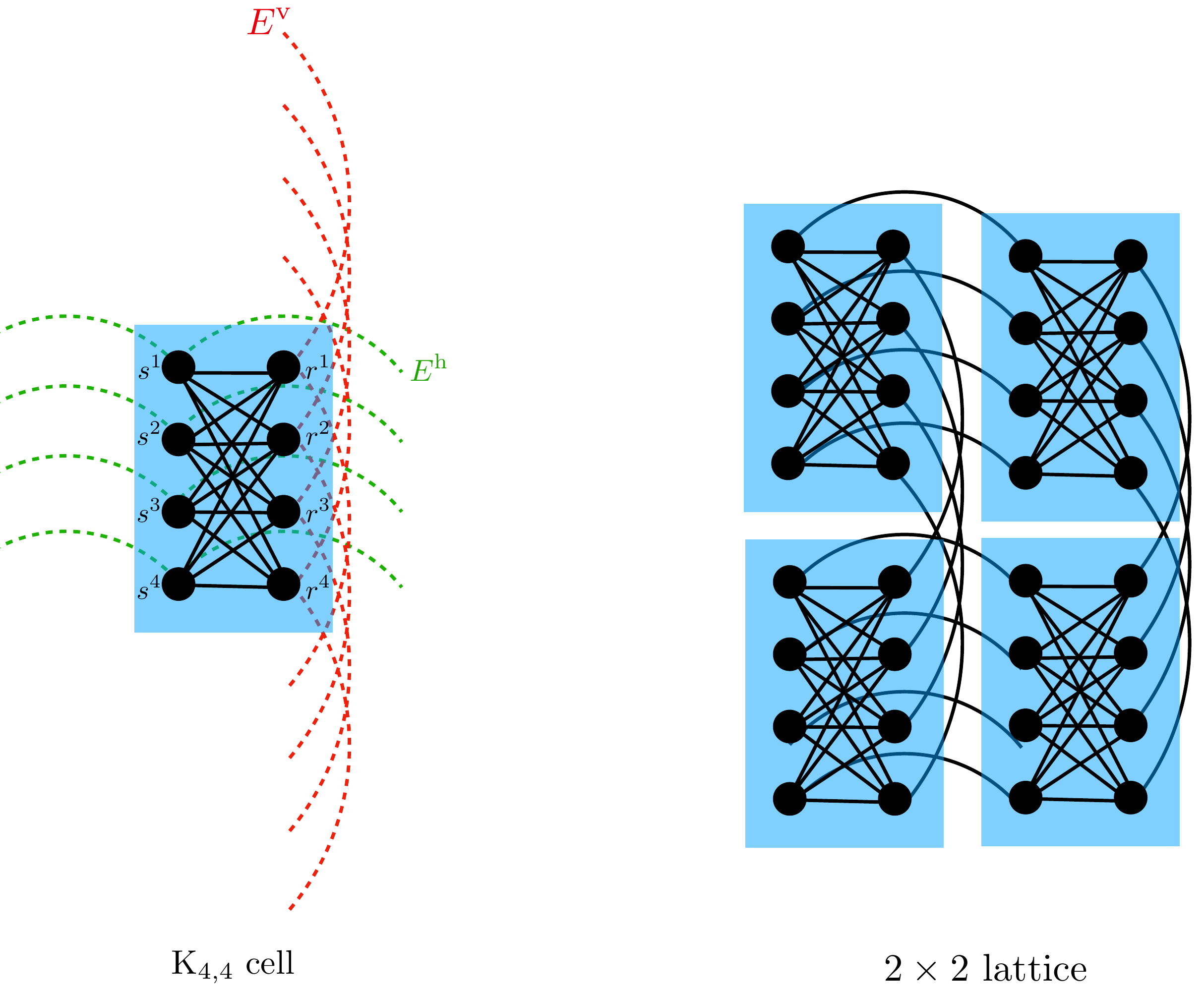}
\caption{$\mathrm{K}_{4,4}$ forms a ``unit cell" of the Chimera graph.  In the left panel, a single cell is shown along with all possible edges to adjacent horizontal and vertical cells.   For later convenience in Section \ref{sec:color}, we also label the 8 vertices in $\mathrm{K}_{4,4}$ in the left panel.   In the right panel, a $2\times 2$ lattice is shown (note that we do not include periodic boundary conditions).}
\label{fig:K44}
\end{figure}

The hardware employed by D-Wave Systems studies QA on a Chimera graph of length $L$.   The Chimera graph can be written in the form (\ref{eq:lattice}), with $(V_{i,j},E_{i,j}) = \mathrm{K}_{4,4}$;   $E_{i,j}^{\mathrm{h}}$ connects the ``left" half of $\mathrm{K}_{4,4}$ between two horizontally adjacent cells;  $E_{i,j}^{\mathrm{v}}$ connects the ``right" half between vertically adjacent cells.  This is best explained with a picture, which we show in Figure \ref{fig:K44}.   For Chimera graphs, $n=8$, $e=16$, $e_{\mathrm{h}} = e_{\mathrm{v}}=4$, and $\bar k \le 6$.

To implement a QUBO defined on graph $G$ as a modified QUBO on a lattice graph $\Lambda$, we must find a minor embedding of $G=(V,E)$ into $\Lambda = (V_\Lambda, E_\Lambda)$.   A minor embedding $\phi:G\rightarrow \Lambda$ is a map with the following properties: (\emph{i})  for each $v\in V$, there exists a connected subgraph $(S_v, T_v) \subset (V_\Lambda,E_\Lambda)$;  (\emph{ii}) the sets $S_v$ are disjoint for all $v\in V$;  (\emph{iii}) for each $(uv)\in E$, there exist vertices $u^\prime \in S_u$ and $v^\prime \in S_v$ with $(u^\prime v^\prime) \in E_\Lambda$.    One conventionally encodes the QUBO (\ref{eq:QUBO}) as \cite{bian} \begin{equation}
H_\Lambda = H(x_{1^\prime},\ldots, x_{N^\prime}) + \alpha \sum_v \sum_{i\ne j \in S_v} \left[ x_i(1-x_j) + x_j(1-x_i) \right],
\end{equation}
where the parameter $\alpha$ is chosen to be sufficiently large, such that $\min(H_\Lambda) = \min(H)$.

\subsection{Embedding a Complete Graph}
There is a general strategy to embed the complete graph $\mathrm{K}_N$, whose edge set consists of all pairs of $N$ vertices, on a lattice $\Lambda$ of length $L$.  Without loss of generality, we choose the vertex set to be $\lbrace 1,\ldots, N\rbrace$.  Let us suppose that $V_{i,j}$ contains at least two vertices $u_{i,j}$ and $v_{i,j}$ obeying $(u_{i,j}v_{i,j})\in E_{i,j}$, $u_{i,j} \in \partial E_{i,j}^{\mathrm{h}}$ and $v_{i,j} \in \partial E^{\mathrm{v}}_{i,j}$.   Then a minor embedding of $\mathrm{K}_N$ into a lattice of length $L=N$ is as follows.  The vertices $i$ are mapped to subgraphs \begin{equation}
S_i = \bigcup_{j=1}^N \lbrace u_{i,j}, v_{j,i} \rbrace.
\end{equation}
Edges are mapped in a straightforward manner.  

\begin{figure}[t]
\centering
\includegraphics[width=2.25in]{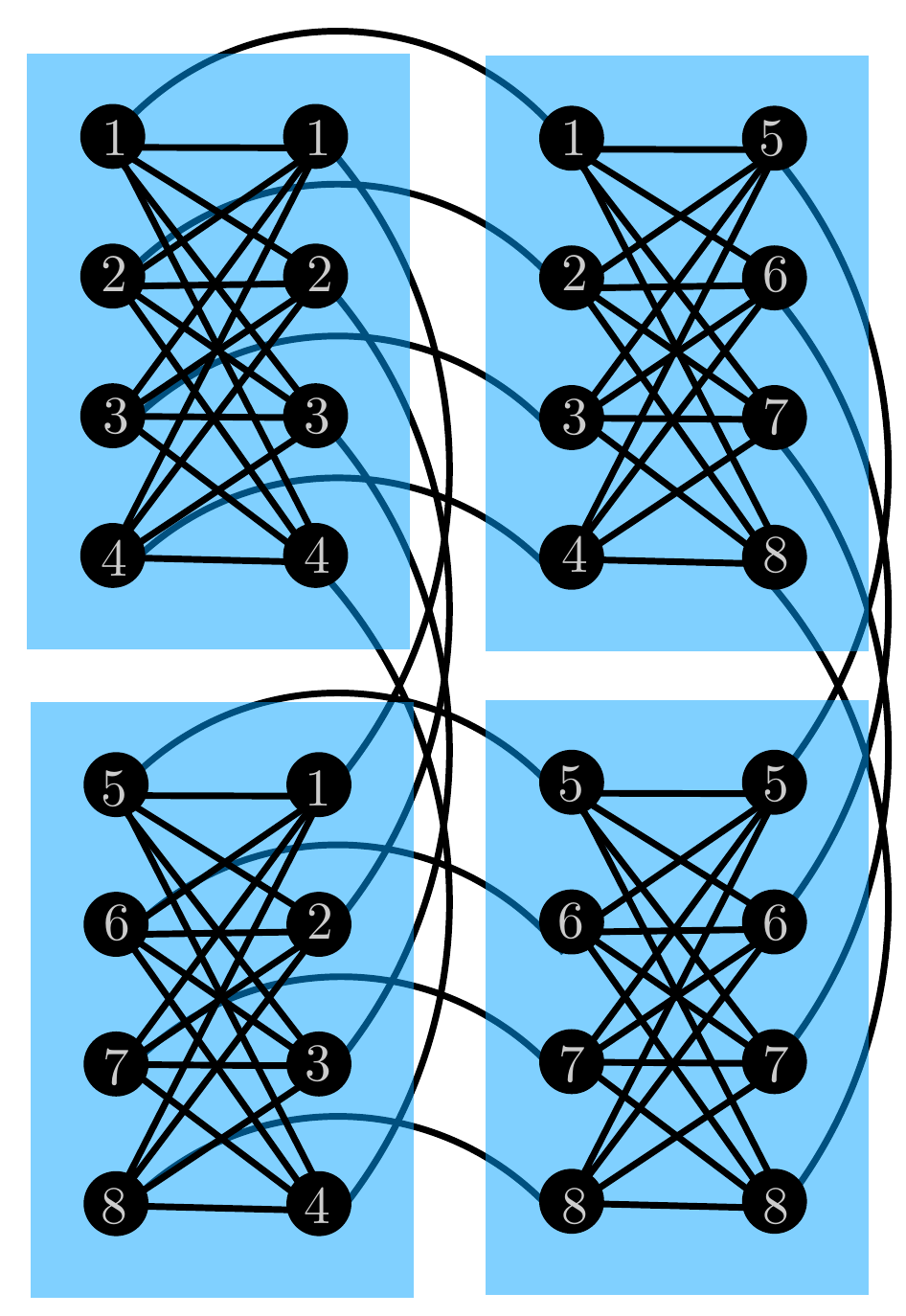}
\caption{Embedding $\mathrm{K}_8$ in a $2\times 2$ Chimera lattice with $\mathrm{K}_{4,4}$ cells.}
\label{fig:K44complete}
\end{figure}

On the Chimera graph, a more efficient embedding can be found \cite{bunyk}.   Using the structure of $\mathrm{K}_{4,4}$, we may embed $\mathrm{K}_N$ on a lattice with $L = \lceil \frac{1}{4}N \rceil$, as shown in Figure \ref{fig:K44complete}.   While further improvements are possible \cite{bian, zaribafiyan}, we will see that the minor embedding of Figure \ref{fig:K44complete} has  useful properties, which we describe in Section \ref{sec:tile}.  A more effective strategy for embedding a single $\mathrm{K}_N$ on the Chimera graph may also be found in \cite{boothby};  it is unclear whether this embedding can be generalized to our later constructions.    We emphasize that the scaling $L \sim N$, and thus $N^\prime \sim N^2$,  holds for generic minor embeddings of $\mathrm{K}_N$.  To see this, note that there are $\frac{1}{2}N(N-1)$ edges in $\mathrm{K}_N$.   The total number of edges in the lattice $|E_\Lambda| \sim L^2 \sim N^\prime$.   Thus $N^\prime \sim N^2$ for any lattice.   

Since no graph of $N$ vertices is harder to embed than $\mathrm{K}_N$,  we conclude that $N \le N^\prime \le nN^2$.    As we discussed in the introduction, quadratic overhead in $N^\prime$ is a serious drawback.  It is important to make $N^\prime$ as small as possible.  One case where it is not possible to asymptotically improve the scaling $N^\prime \propto N^2$ is a random fully connected spin glass with $N^2$ unique coupling constants.   However, for combinatorial problems discussed below, there are a significant number of non-random constraint terms in the QUBO, and we will explicitly construct Hamiltonians with improved asymptotic scaling.

\subsection{Constraints from Spatial Locality} \label{sec:locality}
As we will see, the most serious obstruction to improving the scaling of $N^\prime$ with $N$ is the following problem: consider a connected subgraph $R\subset \Lambda$ of vertices, which consists of $r_1 \times r_2$ cells arranged in a rectangle.   Let $E^\Lambda_{R,R^\mathrm{c}}$ denote edges in $E_\Lambda$ between $R$ and $R^{\mathrm{c}}$.  The number of these edges is \begin{equation}
|E^\Lambda_{R,R^\mathrm{c}}| = (e_{\mathrm{h}}+e_{\mathrm{v}}) (2r_1+2r_2).
\end{equation}
In contrast, the number of vertices in $R$ is \begin{equation}
|R| = nr_1r_2
\end{equation}
and is far larger as $r_{1,2}$ grow large.

Now suppose that we wish to divide up an embedded QUBO into bits that will lie within $R$, and bits that will lie outside of $R$.   Let us divide up the physical vertex set into three disjoint sets $V = V_R \cup V_{R^{\mathrm{c}}} \cup V_{R,R^{\mathrm{c}}}$ that correspond to vertices whose chains lie entirely within $R$, entirely in $R^{\mathrm{c}}$, and in both respectively.   Let $E_{R,R^{\mathrm{c}}} \subset E$ denote the edges connecting $V_R$ to $V_{R^{\mathrm{c}}}$.    Any vertex in $V_{R,R^{\mathrm{c}}}$ necessarily takes up at least one of the edges in $E^\Lambda_{R,R^{\mathrm{c}}}$, as the chain must be propagated along lattice edges.   We conclude that \begin{equation}
|E^\Lambda_{R,R^{\mathrm{c}}}| \ge |V_{R,R^{\mathrm{c}}}| + |E_{R,R^{\mathrm{c}}}|.
\end{equation}
Including vertices in $R$ and $R^{\mathrm{c}}$ -- i.e., propagating long chains -- is extremely costly and greatly increases the size of an embedding into a lattice.   In special (but important)  cases below, we will solve the problem of embeddings where every physical bit embeds in a long chain.

A more serious problem involves the bound $|E^\Lambda_{R,R^{\mathrm{c}}}| \ge |E_{R,R^{\mathrm{c}}}|$.  In many NP-hard combinatorics problems, it is impossible (or, at least, unclear how)  to find new QUBO formulations with sparser interaction graphs.   For example, in the graph coloring problem, a highly connected graph appears to have an unavoidably connected QUBO (see Section \ref{sec:color}).

\subsection{Coupling Constants}
In addition to minor embedding, there is another experimental challenge that we will also address in this paper:  imperfection in the encoding of QUBO parameters $h_{i,j}$, as defined in (\ref{eq:QUBO}).    Let $h^0_{i,j}$ denote the intended couplings in (\ref{eq:QUBO}).  Without loss of generality, we may multiply $H$ by an overall constant prefactor such that \begin{equation}
\left| h_{i,j}\right| \le 1 + \mdelta_{i,j},  \label{eq:couplingrange}
\end{equation}
with \begin{equation}
\mdelta_{i,j} = \left\lbrace \begin{array}{ll} 1 &\  i=j \\ 0 &\ i\ne j\end{array}\right..
\end{equation}
Present day experiments on QA are limited to couplings obeying (\ref{eq:couplingrange}) \cite{bian}.    Furthermore, one cannot  experimentally simulate couplings with arbitrary precision:  the couplings in experiment $h^{\mathrm{exp}}_{ij}$ can be modeled as \cite{bian}
\begin{equation}
 h_{i,j}^{\mathrm{exp}} \approx h_{i,j} + 0.03 \sigma_{i,j} 
\end{equation}
where $\sigma_{i,j}$ are independent, identically distributed, zero-mean, unit-variance Gaussian random variables.   Hence, it is also important to find strategies for eliminating large ranges in the values of $h_{i,j}$.

\section{Unary Constraints}\label{sec:tree}
The following two sections each contain a warm-up problem:  embedding the QUBO for a trivial combinatorics problem onto a lattice.   These simple problems will form the foundation for a new embedding for a non-trivial combinatorial problem in later sections.     

This section addresses embedding unary constraints on $N$ bits:  \begin{equation}
H = \left(1-\sum_{i=1}^N x_i\right)^2.  \label{eq:naiveunary}
\end{equation}
The form (\ref{eq:naiveunary}) requires embedding the complete graph $\mathrm{K}_N$, and so $N^\prime \sim N^2$.   What we now show is that it is possible to encode a unary constraint with $N^\prime \sim N$.   While we focus on (\ref{eq:naiveunary}) in the discussion below, it is also straightforward to generalize to \begin{equation}
H = \left(y-\sum_{i=1}^N x_i\right)^2;
\end{equation}
where $y\in \lbrace 0,1\rbrace$ now allows us to include the possibility that none of the $x_i$ are 1.

\subsection{A Fractal Embedding}
Let $n = \lceil \log_2  N\rceil$.   Now consider the Hamiltonian  \begin{equation}
H = \left(1-\sum_{k=1}^{2^{n-1}} y_k\right)^2 + \sum_{k=1}^{2^{n-1}}\left(y_k - x_{2k}-x_{2k-1}\right)^2 + \sum_{k=N+1}^{2^n} x_k.
\end{equation}
The only ground states of this Hamiltonian are unary ground states, in which exactly one of the $x_1,\ldots, x_N$ is equal to 1.   By introducing the ancilla $y$ variables, we have changed the connectivity of the interaction graph:  now it is more sparse, since the total number of edges is $|E| \approx \frac{1}{2}(\frac{N}{2})^2 + 3\frac{N}{2}$ (the first term comes from the $y$-constraints, and the second from the $xy$-constraints).   For the price of introducing more vertices, we have removed $\frac{3}{4}$ of the  edges (as $N\rightarrow \infty$).   Since our embedding of $\mathrm{K}_N$ into lattice $\Lambda$ has an excess of vertices, this seems to be a welcome tradeoff.

We now recursively continue this approach.   Letting $y_k^j$ denote ancilla variables used to introduce constraints at level $j$  (the Hamiltonian shown above has $y_k = y^{n-1}_k$),  we obtain the following QUBO for encoding a unary constraint: \begin{equation}
H = \left(1-y^1_1 - y^1_2\right)^2 + \sum_{j=1}^{n-2}\sum_{k=1}^{2^j} \left(y^j_k - y^{j+1}_{2k-1} - y^{j+1}_{2k}\right)^2 + \sum_{k=1}^{2^{n-1}}\left(y_k^{n-1} - x_{2k}-x_{2k-1}\right)^2 + \sum_{k=N+1}^{2^n} x_k. \label{eq:2n}
\end{equation}
The number of edges in  this Hamiltonian is \begin{equation}
|E| = 1+  \frac{3}{2}\sum_{j=1}^{n-1} 2^{j} = 1 + 3\left(2^{n-1}-1\right) \le 3N
\end{equation}
while the  number of vertices is \begin{equation}
|V| = \sum_{j=0}^{n-1} 2^{n-j} \le 2^{n+1} \le 4N.
\end{equation}

\begin{figure}[t]
\centering
\includegraphics[width=1.75in]{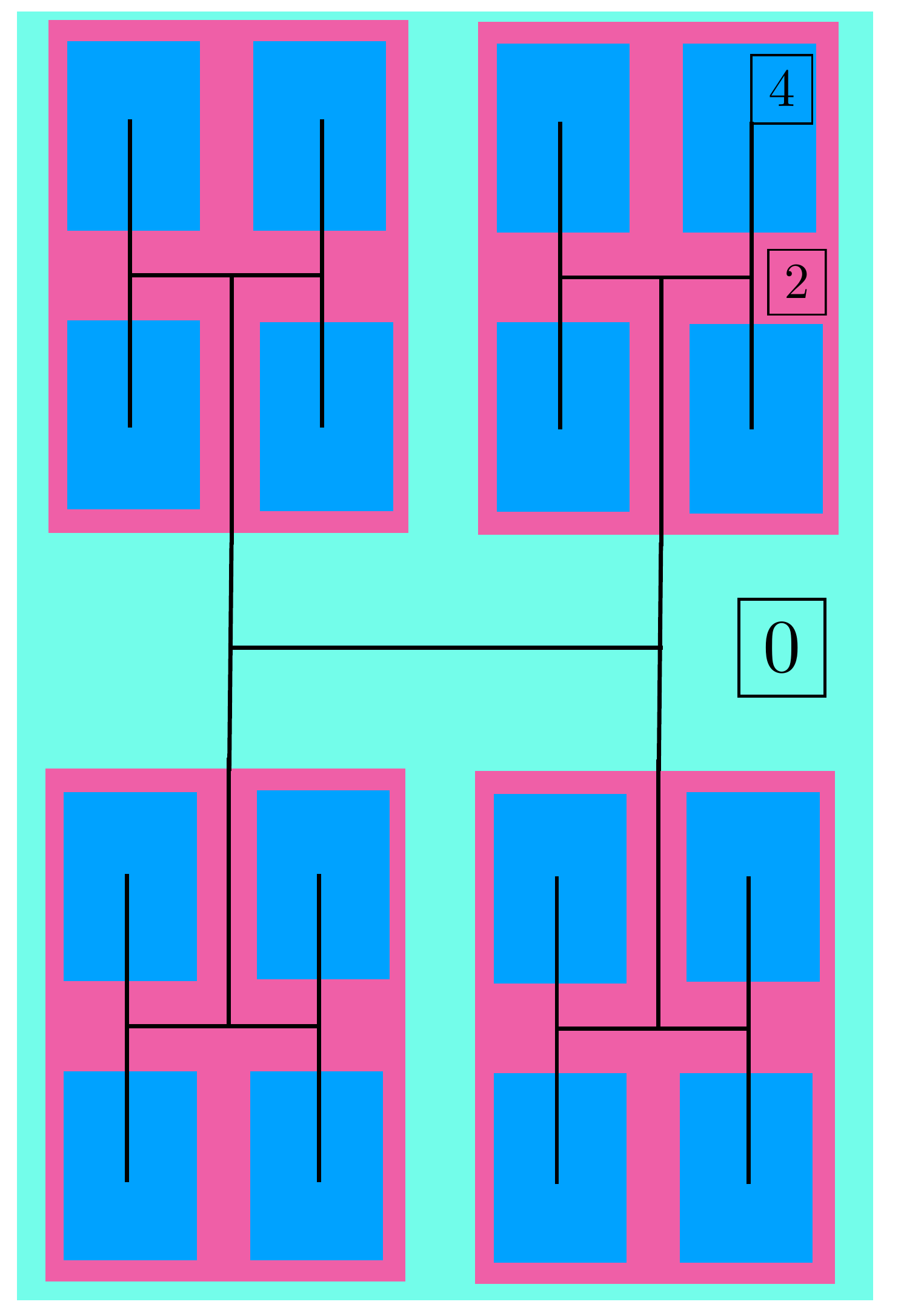}
\caption{The logic of the fractal embedding as a repeatable tiling of the plane (numbers denote the ``layer" of the tree).}
\label{fig:fractal}
\end{figure}

We are now ready to explicitly construct a mapping with $L\sim \sqrt{N}$, and thus $N^\prime \sim N$.    The key observation is that the connectivity graph of the Hamiltonian in (\ref{eq:2n}) is a tree.   Inspiration for how to embed a tree into a lattice can be found in nature:  the circulatory system of animals is approximated with a locally treelike space-filling fractal \cite{west}:   larger veins branch off into smaller veins multiple times before reaching the smallest-scale structures.   Furthermore, this treelike structure is embedded in three space dimensions in a qualitatively efficient manner.    We now mimic this structure when embedding our tree.  It is easiest to describe the fractal mapping of our tree into the lattice with a picture:  see Figure \ref{fig:fractal}.   This embedding could be further improved, but such improvements can only reduce $N^\prime$ by an O(1) factor.   The treelike embedding strategy will also generalize naturally to more complicated problems in Section \ref{sec:NK}.

To explicitly determine $L$, we use the following simple, recursive logic.   As shown in the figure, adding two layers to the tree (increasing  $n$ by 2, or multiplying $N$ by 4) will require approximately doubling the size of the embedding.  More precisely, as hinted at in the right panel of the figure:  \begin{equation}
L_{n+2} = 2L_n + 1,  \label{eq:Lrecursive}
\end{equation}
where $L_n$ denotes the length of the cell graph required to embed a unary constraint with $2^n$ bits.   The $+1$  arises from the additional layer that must be added to allow for ancillas to carry the bit of information from inner layers of the tree towards the `center'.    This is a simple recursive relation which is solved by the substitution \begin{equation}
L_{2m+1} \equiv 2^m K_m.  \label{eq:LmKm}
\end{equation}
Note that $K_0=L_1=1$.   We find \begin{equation}
K_{m+1}-K_m = 2^{-(m+1)},
\end{equation}
which leads to \begin{equation}
K_m-1 =  \sum_{j=1}^m 2^{-j} = 1-2^{-m}.
\end{equation}
Thus, for $n$ an odd integer: \begin{equation}
L_{n} = 2^m (2-2^{-m}) = 2^{(n+1)/2}-1.
\end{equation}
Thus we have  found an  embedding for the unary constraint employing \begin{equation}
L = 2\sqrt{N_*} -1  \label{eq:2sqrtN}
\end{equation}
where $N_*$ is  the number of cells in which the final part of the unary constraint is encoded;  note $N_* \propto N$.   Thus  $N^\prime = \mathrm{O}(N)$ as advertised:  a finite fraction of all vertices in the lattice are being used to encode physical bits $x_i$.

\begin{figure}[t]
\centering
\includegraphics[width=1.75in]{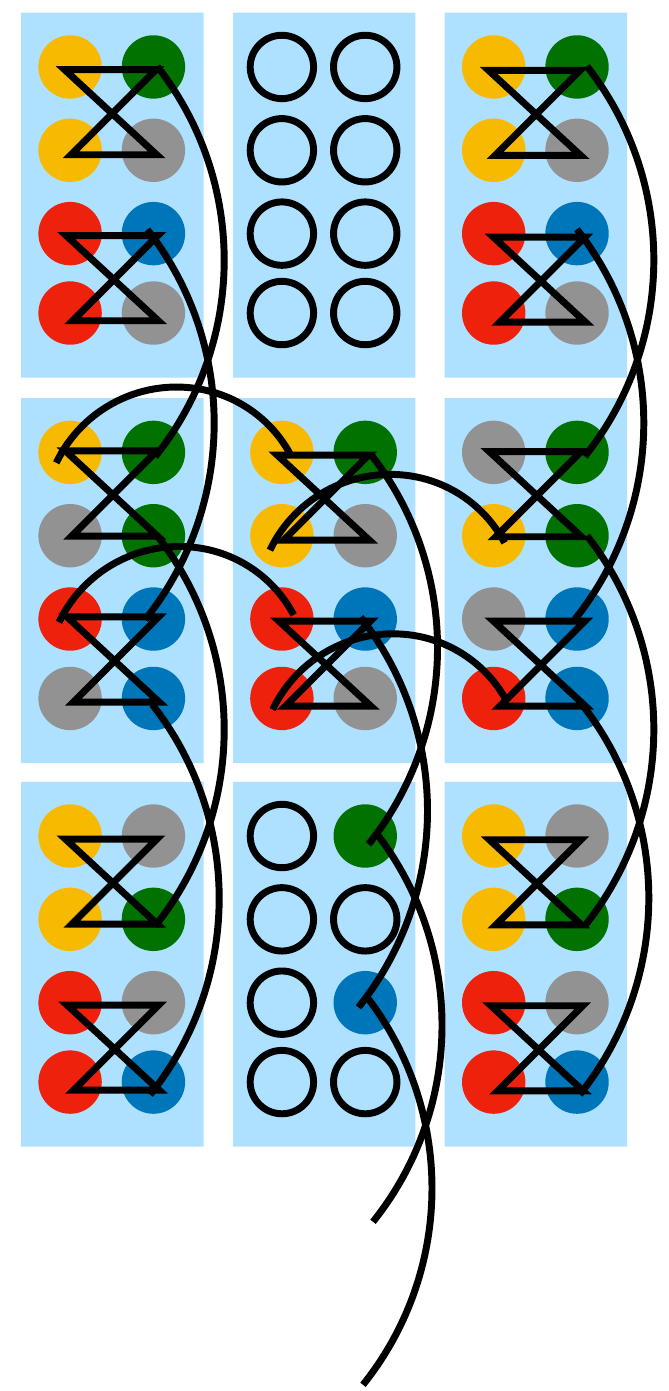}
\hspace{1in}
\includegraphics[width=1.75in]{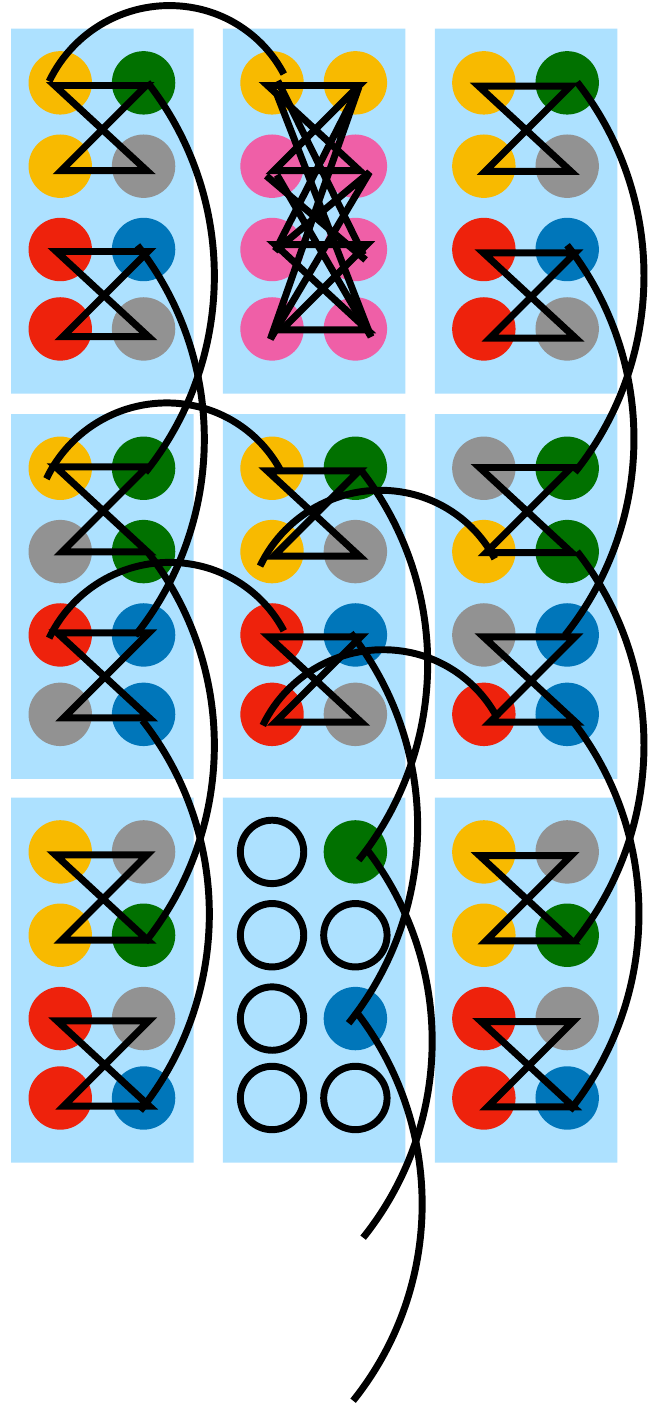}
\caption{Embedding a unary constraint into a Chimera lattice.   Red/yellow and green/blue vertices are used to denote physical bits from (\ref{eq:2n}); gray bits are the auxiliary bits analogous to $s_w$ in (\ref{eq:auxsw}).  The dangling edges at the bottom of the figure can be used to connect this subtree to a larger part of the treelike  constraint embedding.   Left:  the treelike embedding as outlined in Figure \ref{fig:fractal}, which leads to (\ref{eq:2sqrtN}).   Right:  the improved embedding described in Section \ref{sec:treeopt}, with $J=4$, in which ``gaps" in the embedding are filled with more bits.   Pink bits denote the 3 additional bits that have been added to the unary constraint.}
\label{fig:chimunary}
\end{figure}

In the chimera lattice, it is possible to obtain $N_* = \frac{1}{4}N$;  here $N_*$ is the number of $\mathrm{K}_{4,4}$ cells which form the leaves of the tree, and $N$ is the number of variables in the constraint.   To achieve this value of $N_*$ we must  encode two subtrees simultaneously in each cell.   We replace an intermediate constraint in the tree  $H=(z-x-y)^2$ (here $z,x,y$ denote bits) as follows: let $s_z,s_w$ and $s_x,s_y$ denote spins on each half of $\mathrm{K}_{2,2}$ respectively.   Then the ground states of $(s_z - s_x - s_y - 1)^2$ are the same as the ground states of 
\begin{equation}
H  =  s_x + s_y - s_z(1+s_x+s_y) + s_w(s_x-s_y)   \label{eq:auxsw}
\end{equation}
up to an additive (and unimportant) constant.   Since we can embed each intermediate step in the tree in $\mathrm{K}_{2,2}$, it is possible to embed a constraint with $N_* = \frac{1}{4}N$.   See Figure \ref{fig:chimunary} (left panel) for an explicit picture of how (part of) such an embedding goes.  Using (\ref{eq:2sqrtN}) we obtain $L=\sqrt{N}-1$ for the fractal embedding.    

It is also possible to embed a complete graph using $L= \frac{1}{4}N$.   It is rapidly advantageous to use the fractal embedding.  For example, consider trying to embed a unary constraint on $N=16$ variables.   Using the complete graph embedding this requires $L=4$,  but using the fractal embedding we can achieve this using $L=3$.    Using a $L=15$ sublattice of Chimera, we may embed a unary constraint on 256 bits.  Using the square embedding for the complete graph described above,  the maximal number of bits embeddable with present day hardware is 64, using the full $L=16$ lattice.

More generally, suppose that we have a $\mathrm{K}_{J,J}$ lattice (the previous two paragraphs discuss $J=4$).   Without filling in the gaps in the embedding, we find from (\ref{eq:2sqrtN}) that we can embed a unary constraint on $N$ variables in a lattice of length  \begin{equation}
L = 2\sqrt{\frac{N}{J}} - 1.
\end{equation}
In contrast, using the complete graph embedding we find $L=N/J$.   As before, we find that the new treelike embedding become competitive with existing methods once $L=3$, as we can embed a constraint of $N=4J$ variables, whereas the complete graph embedding requires $L=4$ for this problem size.

\subsection{Optimization} \label{sec:treeopt}
From the left panel in Figure \ref{fig:chimunary}, it is clear that there is some ``unused space" in our treelike embedding of constraints.    The right panel of Figure \ref{fig:chimunary} suggests how this can be fixed -- simply add some further branches to the tree that fill in the unused space.  For example, in the Chimera  graph, filling in the unused cells at the deepest levels of the tree allows us to add an extra 3 bits for every 16 we started with.   This is an O(1) improvement in the embedding size, but is still helpful.

\begin{figure}[t]
\centering
\includegraphics[width=3in]{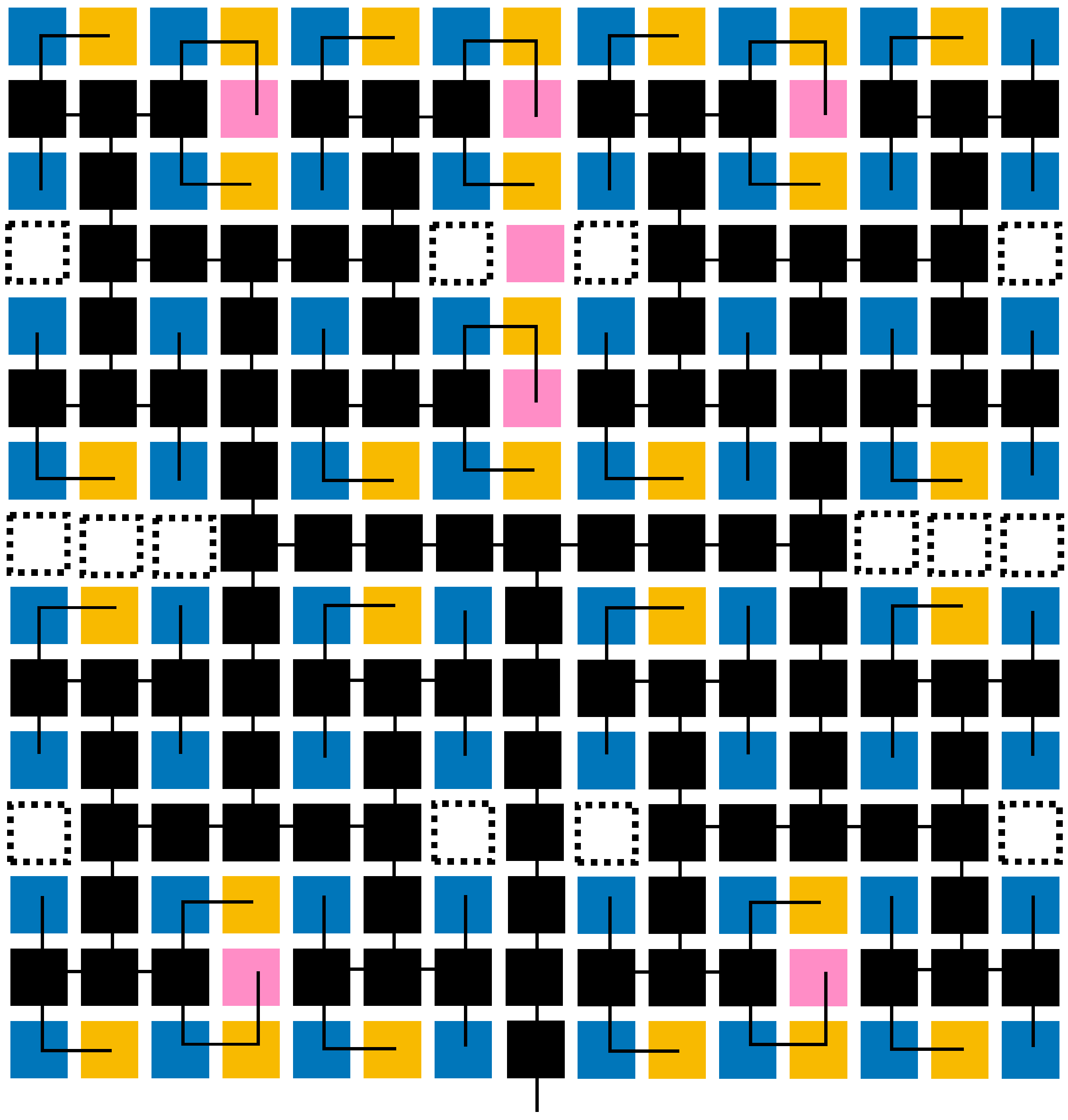}
\caption{A sketch of how to fill in the treelike embedding of a unary constraint, thus increasing the value of $N$ which can be embedded in a given size lattice.   Black squares denote cells which are required to encode higher levels in the tree; blue squares denote the original leaves of the tree, and yellow/pink squares denote the added cells which allow us to increase $N$.  White squares with dashed boundaries denote unused tiles. }
\label{fig:filltree}
\end{figure}

Filling in the tiles as in the right panel of Figure \ref{fig:chimunary} gives us an extra $J-2$ bits to work with.  One bit in an existing cell is replaced with a variable that is fixed to be sum of a subset (not an independent bit), and that this new variable must have an ancilla in the new cell (as the unused cells are connected to the used cells by only a single edge).     At higher levels of the tree, we can repeat the same ideas.      As in Figure \ref{fig:chimunary}, at the deepest level of the tree, we have embedded the constraint such that any extra branches must be included in horizontally adjacent tiles.   So Figure \ref{fig:filltree} shows one way to partially fill in the tree in a $15\times 15$ lattice, adding extra branches to the tree (yellow cells) for every tile horizontally adjacent  to one of our original leaves (blue cells).   As the extra branches have encoded a complete graph, we may also freely  add a few more bits in empty cells adjacent to the yellow cells: these are now pink. 

Let us now count how many additional bits we can encode in a unary constraint.   Every time we add an additional branch, we can add $J-2$ additional bits.   We conclude that (\ref{eq:Lrecursive}) generalizes to \begin{subequations}\begin{align}
L_{m+1} &= 2L_{m}+1, \\
N_{m+1} &= 4N_m + (J-2)L_m
\end{align}\end{subequations}
with $L_1=1$, $N_1 = J$.   We find that \begin{equation}
N_m = 4^m \left[\frac{J}{4}+(J-2)\sum_{\ell=2}^m \left(\frac{1}{2^{\ell+1}} - \frac{1}{4^\ell}\right) \right].
\end{equation}
At large $L$ and $N$, we conclude that \begin{equation}
L \approx \sqrt{\frac{12N}{5J-4}},
\end{equation}
which is approximately 22\% smaller than (\ref{eq:2sqrtN}) at large $J$ and $N$.   Indeed, filling in the unused cells in the lattice could only lead to an O(1) improvement, because $L=\mathrm{O}(\sqrt{N})$ is mathematically optimal.

We have belabored the embedding of a trivial unary constraint problem because the general philosophy of breaking up problems into smaller subproblems and embedding the subproblems effectively will prove quite useful in the rest of this paper.  In particular, the treelike embedding is especially helpful in the context of Section \ref{sec:NK}, where the problem size becomes ``larger" at each step of the tree.   The treelike construction then minimizes the number of bits used in embedding ``large" problems.

\section{Adding in Binary}\label{sec:binary}
A second warm-up problem involves the efficient encoding of the addition of two large integers.  In this section, we will focus on finding a QUBO with small coupling constants, and will not address the embeddability of the QUBO.  

An ineffective (but standard \cite{lucas}) way of adding two integers is as follows. Let  $X_1$ and $X_2$ be two integers obeying $1\le X_1,X_2 \le 2^n-1$.   Their sum is then bounded by $2^{n+1}-1$.   Writing $Y=X_1+X_2$, and \begin{equation}
X_i =  \sum_{j=0}^{n-1} 2^j x_i^j, \;\;\;\;\;\; Y = \sum_{j=0}^{n} 2^j y^j,
\end{equation}
the simple Hamiltonian \begin{equation}
H = \left(2^n y^n + \sum_{j=0}^{n-1}2^j (y^j - x^j_1-x^j_2)\right)^2   \label{eq:sec2first}
\end{equation}
enforces that $Y=X_1+X_2$.    For the D-Wave device, (\ref{eq:sec2first}) involves unacceptably large coupling constants when $n\gtrsim 4$.

A better approach to adding two integers is to encode the elementary school addition algorithm as a QUBO.  Let us review, in symbols, this algorithm.  We first add together the last two digits of $X_1$ and $X_2$:  
\begin{equation}
y^0 = (x_1^0 + x_2^0)\; (\mathrm{mod}  \; 2). 
\end{equation} 
If $x_1^0+x_2^0 = 2$, then we must carry a 1 in to the next column of digits.  So let us define the  integer \begin{equation}
z^1 =  \frac{x_1^0 + x_2^0 - y^0}{2}.
\end{equation} Then
 \begin{equation}
y^1 = (x_1^1 + x_2^1 + z^1)\; (\mathrm{mod} \; 2). 
\end{equation}
This process repeats in a straightforward way:
\begin{equation}
y^j = (x_1^j + x_2^j + z^j)\; (\mathrm{mod} \; 2) \label{eq:yj}
\end{equation}
and \begin{equation}
z^{j+1} =  \frac{x_1^j + x_2^j + z^j - y^j}{2}  \label{eq:zj}
\end{equation} (with $z^0=0$).   When we reach $j=n-1$, we simply set $y^{n}=z^{n}$.   

We can directly convert this into a QUBO: $x_1^j$, $x_2^j$, $y^j$ and $z^j$ are all binary bits, so consider the quadratic Hamiltonian \begin{equation}
H_{\mathrm{binary}}[y^j, x_1^j, x_2^j; z^j] =  \sum_{j=0}^{n-1} \left(y^j  +2z^{j+1} - x_1^j - x_2^j- z^j\right)^2. \label{eq:binary}
\end{equation}
Instead of using $3n+1$ bits, as in (\ref{eq:sec2first}), we are now using $4n$ bits (recall that $y^n=z^n$ and $z^0=0$).   The term in parentheses can be zero (for binary variables) if and only if (\ref{eq:yj}) and (\ref{eq:zj}) are both obeyed for all $j$.   All of the coupling constants in $H_{\mathrm{binary}}$ are now finite and independent of $n$.  We conclude that this embedding is optimal (at least,  up to a constant prefactor).  Importantly, $x_1^j$ and $x_2^j$ may themselves be treated as bits in the above expression, and it remains quadratic.

\section{Number Partitioning and Knapsack} \label{sec:NK}
Let us now discuss how to effectively embed some simple NP-hard combinatorics problems.
\subsection{Number Partitioning}\label{sec:numpart}
We begin with the NP-complete version of the number partitioning problem:  given a list of integers $\lbrace n_1,\ldots ,n_N\rbrace$ obeying $1\le n_i \le 2^M-1$, is it possible to divide the integers into two sets such that the sum of integers in both sets is the same?   In other words, if $s_i = \pm 1$ is a binary spin variable used to assign integer $n_i$ to one of two sets, does the Hamiltonian \begin{equation}
H = \left(\sum_{i=1}^N n_i s_i \right)^2 
\end{equation}
have a ground state energy of 0?

Clearly, the above Hamiltonian suffers from two major drawbacks:   the coupling constants in $H$ could range from 1 to $2^{2M}$ in size.   The number partitioning problem is most challenging when $N\sim M$ \cite{fontanari, mertens}, so this is a serious problem.   Secondly, the QUBO above is all-to-all connected, meaning that $N^\prime \sim N^2$.    

By combining the methods of Sections \ref{sec:tree} and \ref{sec:binary}, we can solve both of these problems.   Let us denote \begin{equation}
W  = \frac{1}{2}\sum_{i=1}^N n_i.
\end{equation}
We assume $W$ is an integer.   Our goal will be to add together a large number of integers:  at each step, the addition is performed using the algorithm of Section \ref{sec:binary} -- the additions are arranged in the space-filling fractal structure of Section \ref{sec:tree} to exploit a far larger fraction of all vertices in the lattice for physical bits.   The structure of the QUBO will be as follows.   Let \begin{equation}
m = \lceil \log N \rceil.   
\end{equation}
In this paper, the logarithm is always understood to be base 2.   Let $\overline{\mathbb{I}}[\cdots]$ denote the (backwards) indicator function, which equals 0 if its argument is true, and 1 if its argument is false.     Then we wish to write \begin{equation}
H = \overline{\mathbb{I}}\left[X^1_0 = W \right] + \sum_{j=0}^{m-2} \sum_{k=1}^{2^j} \overline{\mathbb{I}}\left[X^k_j = X^{2k}_{j+1}+X^{2k-1}_{j+1}\right] + \sum_{k=1}^{2^{m-1}} \overline{\mathbb{I}}\left[X^k_{m-1} = n_{2k-1}x_{2k-1}+n_{2k}x_{2k}\right]  \label{eq:HNP}
\end{equation}
in QUBO form.   The auxiliary variables $X^k_j$ encode a pairwise summation of the integers $n_i$.   So ultimately, we will plan to arrange the bits in the lattice graph in the fractal structure of Section \ref{sec:tree}.

We now address how to implement each indicator function.  We start with the third term in (\ref{eq:HNP}).  This is accomplished as follows:  we use the binary addition algorithm of Section  \ref{sec:binary}, with the added constraint that the two integers being added are either 0 or the appropriate $n_i$.    Let \begin{equation}
n_i = \sum_{j=0}^{M-1} n_i^j 2^j, \;\;\;\; X_i^k = \sum_{j=0}^{M-1+m-k} X^{k,j}_i 2^j,
\end{equation}with $n_i^j \in \lbrace 0,1\rbrace$.   To make sure that we add either 0 or $n_i$, we simply write
 \begin{equation}
\overline{\mathbb{I}}\left[X^k_{m-1} = n_{2k-1}x_{2k-1}+n_{2k}x_{2k}\right] = \sum_{j=0}^{M-1}\left(X^{k,j}_{m-1} + 2Z^{k,j+1}_{m-1} - Z^{k,j}_{m-1} - n_{2k-1}^jx_{2k-1} - n_{2k}^j x_{2k}\right)^2 .
\end{equation}
The $Z^{k,j}_i$ bits introduced above are the auxiliary bits introduced in Section \ref{sec:binary}.   Because we are adding together two fixed integers, a single bit $x_i$, together with the knowledge of $n_i^j$ and the output/auxiliary bits $X^{k,j}_{m-1}$/$Z^{k,j}_{m-1}$, is sufficient to encode the addition problem.   The number of bits required is $2M+4$.   Keep in mind that there are $\mathrm{O}(N)$ such additions.

The second indicator in (\ref{eq:HNP}) is straightforward:  \begin{equation}
\overline{\mathbb{I}}\left[X^k_j = X^{2k}_{j+1}+X^{2k-1}_{j+1}\right] = H_{\mathrm{binary}}\left[X_j^{k,l}, X_{j+1}^{2k,l}, X_{j+1}^{2k-1,l};  Z^{k,l}_j\right].
\end{equation}
Writing \begin{equation}
W = \sum_{j=0}^{M+m-1} W^j 2^j,
\end{equation}
with $W^j \in \lbrace0,1\rbrace$, the  first indicator in (\ref{eq:HNP}) is simply \begin{equation}
\overline{\mathbb{I}}\left[X^1_0 = W \right] = \sum_{j=0}^{M+m-1} \left(W^j - X_0^{1,j}\right)^2.
\end{equation}

Thus we have found our overall Hamiltonian.  All of the coupling constants are of the same order.  The last step is to explicitly bound $L$, which we do following (\ref{eq:Lrecursive}).   However, we must be aware of  the following subtlety:  each time we move up one layer in the tree, we are adding together integers that grow larger and larger.   As in Figure \ref{fig:fractal}, a doubling of $L$ will approximately allow us to fit a tree with 2 more levels.   Adding 4 $p$-bit integers together, the output may need to be encoded in a $p+2$-bit integer.   At the lowest level of the tree, we are adding together two $M$-bit integers.   Recalling the number of bits necessary to perform binary addition, which was discussed below (\ref{eq:binary}), we conclude that the generalization of (\ref{eq:Lrecursive}) to our embedding of the number partitioning problem in a $\mathrm{K}_{J,J}$ lattice is \begin{equation}
L_{j+2} =  2L_j  + \frac{4(M+j)}{J}.
\end{equation}
We are defining $j=1$ here as the deepest level of the tree, where the specific integers are encoded with the bits $x_i$.    Using (\ref{eq:LmKm}), we find that \begin{equation}
K_j - (2M+4) = \frac{1}{J} \sum_{l=1}^j 2^{2-l}(M+2l) < \frac{4\left[M +4\right]}{J}
\end{equation}
As $2^m \ge N$ and (\ref{eq:2sqrtN}), we conclude that the final lattice length $L$ necessary to encode the full knapsack problem is  \begin{equation}
L \le \frac{12M+40}{J}\sqrt{N} . \label{eq:LfirstNP}
\end{equation}
Assuming $M$ is fixed, we have obtained optimal scaling of $L\sim \sqrt{N}$ instead of $L\sim N$.   However, we also note that the constant coefficients in (\ref{eq:LfirstNP}) are large.  A more detailed study of specific lattice graphs is necessary to understand whether important reductions in such constant prefactors are possible.

One improvement that may not be possible is a parametric improvement of the scaling $L\sim M$.  The reason for this limitation is that we must transmit the integers $X^k_j$ through the tree.  As discussed in Section \ref{sec:locality}, to embed a chain of $M$ bits through a lattice requires a length of $\mathrm{O}(M)$.    For the most difficult problems, where $M\sim N$ \cite{fontanari, mertens}, we obtain $L\sim N^{3/2}$, which is still a parametric improvement over $L\sim N^2$.

Since the algorithm presented in \cite{lucas} requires extremely large coupling constants that cannot be realized experimentally, let us estimate the size of an embedding for the number partitioning problem found by adding the integers pairwise, but not using a treelike embedding.   On an $L\times L$ Chimera lattice, we can embed such a problem so long as \begin{align}
L &< \frac{1}{J}\left[NM + 2\times \left(\frac{N}{2}(M+1) + \frac{N}{4}(M+2) + \cdots\right)\right] \notag \\
&= \frac{1}{J}\left[ -NM + 2\sum_{n=0}^j (M+j)\frac{N}{2^j}\right] < \frac{7NM}{J}.  \label{eq:LsecondNP}
\end{align}
The first factor on the right hand side counts the bits $x_i$ that denote whether a number is included in a given partition; the second factor estimates the number of auxiliary bits $X^{k,l}_j$ and $Z^{k,l}_j$.    Comparing (\ref{eq:LfirstNP}) and (\ref{eq:LsecondNP}), we find that for $J=4$, $M=2$ and $N=16$,  the treelike embedding will be comparable in size to the conventional embedding, based on a complete graph, and both embeddings will require $L\sim 60$.   This may be achievable with hardware advances in the coming years.  Beyond this value of $L$, the treelike embedding becomes more efficient.

\subsection{Knapsack}
Another problem that may be similarly embedded is the  knapsack problem:   given a list of $N$  items, each one with value $V_i$ and weight $W_i$ ($i\in \lbrace 1, \ldots, N\rbrace$), and defining \begin{equation}
W_{\mathrm{tot}} = \sum_{i=1}^N W_i x_i, \;\;\;\;\; V_{\mathrm{tot}} = \sum_{i=1}^N V_i x_i, 
\end{equation} for $x_i \in \lbrace 0,1\rbrace$, find the maximal value of $V_{\mathrm{tot}}$ such that $W_{\mathrm{tot}} \le W_{\mathrm{max}}$.   We assume that all $W_i$ and $V_i$ are integers.

A simple Hamiltonian for the knapsack problem was presented in \cite{lucas}.   Let $m=\lceil \log_2 W_{\mathrm{max}}\rceil$; then  \begin{equation}
H = A \left(\sum_{j=0}^{m-1}2^j y_j + (W_{\mathrm{max}}+1-2^m)y_m - \sum_{i=1}^N W_i x_i \right)^2 - \sum_{i=1}^N V_i x_i   \label{eq:origknapsack}
\end{equation}
with $A$ a sufficiently large constant (e.g., $A= N\max(V_i)$).   The $y_j$ variables are used to encode the constraint that the weight of the items can only be so large, and the constraint enforces that the weight of the included objects is precisely given by a positive integer $\le W_{\mathrm{max}}$.   The second term simply optimizes over the value of the included  objects.   The ground state energy of  this Hamiltonian is (up to a minus sign) the solution to the optimization problem.  As before, we can use indicator functions to write \begin{equation}
H = A \overline{\mathbb{I}}\left[W_{\mathrm{max}}\ge \sum_{i=1}^N W_i x_i \right] - \sum_{i=1}^N V_i x_i.\label{eq:indfunc}
\end{equation}

Unfortunately, the couplings in (\ref{eq:origknapsack}) are very large, and the interactions are all-to-all.   As in the number partitioning problem, we can essentially solve the problem of large coupling constants, and significantly improve upon the connectivity challenge by embedding an equivalent QUBO formulation with a sparser interaction graph.  Let $n=\lceil \log_2  N\rceil$, and let $\ell_*$ be an integer that will encode a ``target" value for $\sum V_i x_i$. The abstract Hamiltonian that we will formulate as a sparse QUBO is  \begin{align}
H &= \overline{\mathbb{I}}\left[ W_{\mathrm{max}} \ge W_{0,1} \right]  + \overline{\mathbb{I}}\left[ 2^{\ell_*+1} > V_{0,1} \ge 2^{\ell_*} \right]  \notag \\
&+ \sum_{j=0}^{n-2} \sum_{k=1}^{2^j} \left(  \overline{\mathbb{I}}[V_{j,k} = V_{j+1,2k-1} + V_{j+1,2k}] + \overline{\mathbb{I}}[W_{j,k} = W_{j+1,2k-1} + W_{j+1,2k}]  \right) \notag \\
&+  \sum_{k=1}^{2^{n-1}} \left(  \overline{\mathbb{I}}[V_{n-1,k} = V_{2k-1}x_{2k-1} + V_{2k}x_{2k}] + \overline{\mathbb{I}}[W_{n-1,k} = W_{2k-1}x_{2k-1} + W_{2k}x_{2k}]  \right).  \label{eq:Hknapsack}
\end{align}
The first line encodes inequality constraints:  the weight in our knapsack is below the maximal value, and the total value of goods stored is within a factor of 2 of $2^{\ell_*}$.  The second line encodes a treelike summation of the values/weights of the goods stored in our knapsack,  which are themselves encoded in the third line.    Because of our constraint that we can only check whether the value of goods stored in the knapsack is within a given range, we must solve the QUBO multiple times for different choices of $\ell_*$ to fully solve the NP-hard optimization problem.   But this will only increase the runtime of the algorithm by a factor  of $\log(N\max(V_i))$, in the worst case.

For simplicity, we first suppose that $W_{\mathrm{max}} = 2^m-1$, that $m>\ell_*>n$, and that $\max(W_i) \le 2^{m^\prime}-1$, $\max(V_i) \le 2^{\ell^\prime}-1$.  Then, the constraint $W_{\mathrm{max}} \ge W_{0,1}$ is trivially encoded by the statement that the total weight that we added up can be encoded as \begin{equation}
W_{0,1} = \sum_{j=0}^{m-1} 2^j w_{0,1}^j.
\end{equation}
 $w_{0,1}^j$ are bits that encode the integer $W_{0,1}$ (as usual).  If the weights are too large for this to be satisfied, then some of the constraints in the second line of (\ref{eq:Hknapsack}) cannot be satisfied.   
 In what follows, we will define the variables $v_{j,k}^\ell$ and $w_{j,k}^\ell$ analogously to above:
  \begin{equation}
V_{j,k} = \sum_{p=0}^{\max(\ell_*,  \ell^\prime + n-k)} 2^p v_{j,k}^p, \;\;\;\;\;\;\;\;\; W_{j,k} = \sum_{p=0}^{\max(m,  m^\prime + n-k)} 2^p w_{j,k}^p.
\end{equation}   
The second constraint on the first line of (\ref{eq:Hknapsack}) simply amounts to $v_{0,1}^{\ell_*} = 1$, which can be easily  implemented by deleting  one bit from the Hamiltonian and replacing its couplings to other bits with suitable single-bit terms.

The Hamiltonians on the second and third  line are encoded exactly as in Sections \ref{sec:binary} and  \ref{sec:numpart}.  They are embedded (and $V_{j,k}$ and $W_{j,k}$ are transmitted via ancilla vertices) via the space-filling fractal pattern of Section \ref{sec:tree}.   Using the same logic as in Section \ref{sec:numpart}, the length of the cell graph  required to encode this Hamiltonian obeys  the recursive relation \begin{equation}
L_{n+2} \le 2L_n + \left[4(n+1+\ell^\prime) + 4(n+1+m^\prime) \right].
\end{equation}  
The constant factor here, in square brackets, arises from the fact that we need to transmit a pair of integers $V_{j,k}$ and $W_{j,k}$;  the parentheses correspond to the terms used to transmit $V$ and $W$, respectively. Using the  same technique as in (\ref{eq:LmKm}), and defining $n_0 = (n+1)/2$, we find \begin{equation}
JL_n = 2^{n_0} \left(1 + \sum_{j=1}^{n_0} (8+8j + 4\ell^\prime + 4m^\prime) 2^{-j} \right) \le 2^{(n+1)/2} \left(25+4\ell^\prime + 4m^\prime  \right).
\end{equation}
In terms of more useful variables:  \begin{equation}
L \le \frac{\sqrt{N}}{J} \left(50 + 8\log \max(V_i) + 8\log \max (W_i)\right).
\end{equation}
Similar to the number partitioning problem, if $V_i$ and $W_i$ are independent of $N$, then we have found (up  to a constant prefactor) an optimal embedding of the knapsack problem, as $L\sim \sqrt{N}$.  However, we expect that the most challenging instances of the knapsack problem (which can be solved in pseudo-polynomial time via dynamic programming \cite{pisinger}) exhibit $\log \max(W_i) \sim \log \max(V_i) \sim N$, in which case we find $L\sim N^{3/2}$, as  in the number partitioning problem.    On such instances, dynamic programming is no longer effective.   There may be other families of knapsack problems that are challenging to solve, yet whose $W_i$ are not very large \cite{pisinger};  at least asymptotically, quantum annealing may be a promising approach for these instances.

\section{Combinatorial Problems on Graphs} \label{sec:combgraphs}
We now turn to the second half of the paper, which focuses on how to embed combinatorial problems on graphs $G=(V,E)$.   Roughly speaking, these typically involve placing some number of bits on every vertex of the graph, and subsequently demanding that various constraints hold whenever two vertices are (or are not) connected by an edge.    Two examples that we will study in some depth in later sections are graph coloring and Hamiltonian cycles.

\subsection{Tileable Embeddings}\label{sec:tile}

Our strategy for embedding combinatorial problems on graphs involves finding a ``tileable embedding."   Namely, we will aim to write the Hamiltonian for the combinatorial problem as a sum of terms \begin{equation}
H = \sum_{v\in V} H_v + \sum_{uv\in E} H_{uv},   \label{eq:tileableH}
\end{equation}
and look for embeddings of the building block Hamiltonians $H_v$ and $H_{uv}$ into small subregions of the lattice graph.   We will subsequently stitch together the embeddings of the small problems to find our final embedding.    For simplicity, let us make the following assumptions.  (\emph{i}) $H_v$ can be embedded in a $\ell \times \ell$ sublattice.     We will refer to these $\ell \times \ell$ subgraphs of the lattice graph as tiles.    (\emph{ii})  $H_u + H_v + H_{uv}$ can be encoded in a $2\ell \times \ell$ and/or $\ell \times 2\ell$ sublattice.  In the former case, the left half of vertices encode $H_u$ and the right half encode $H_v$; in the latter case the top half encodes $H_u$ and the bottom half encodes $H_v$;  in either case, couplings between the two halves encode $H_{uv}$.   We also assume $u$ and $v$ may be swapped, and that $H_{uv}$ has no preferred orientation.  (\emph{iii}) It is possible to use an $\ell\times \ell$ tile to encode nonplanarity.   This is always possible so long as the unit cell of the lattice has at least two vertices and $|\partial E^{\mathrm{h}}_{i,j} \cup \partial E^{\mathrm{v}}_{i,j}| > 1$.  

Assumption (\emph{ii}) above is actually somewhat severe.   Depending on how many couplings are contained in $H_{uv}$, $\ell$ can become rather large.   However, as discussed in Section \ref{sec:locality},  it is not easy to improve upon this constraint without finding fundamentally new embedding strategies.   The embedding strategies which we discuss below are nevertheless state-of-the-art and, in some instances, provide parametric improvements over existing methods.

\begin{figure}[t]
\centering
\includegraphics[width=5.5in]{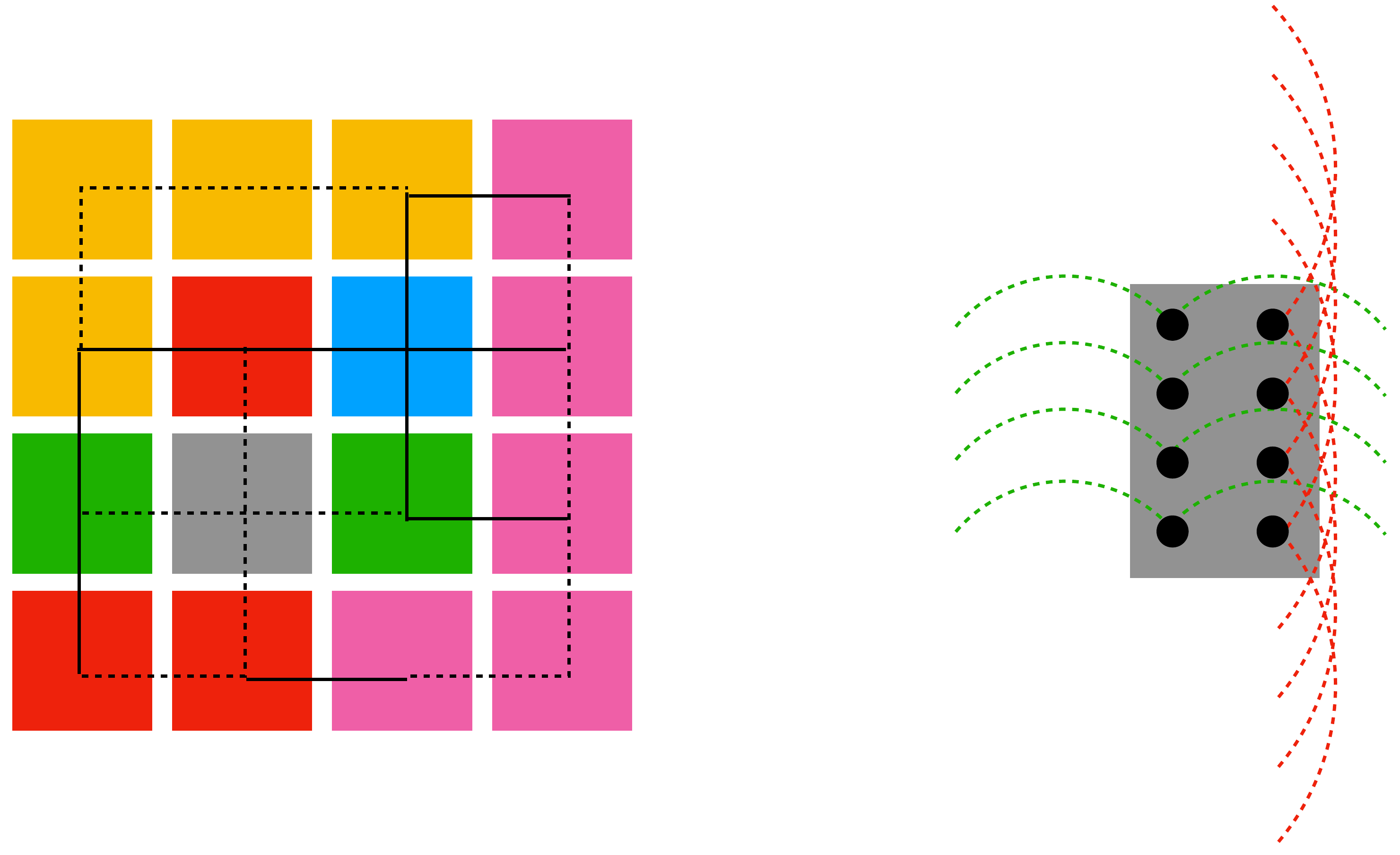}
\caption{The left panel shows a tileable embedding of $\mathrm{K}_5$ in a $4\times 4$ lattice graph.   Each color denotes a tile associated to a different vertex, except for the gray tile which denotes a crossing tile. Dashed lines denote propagation of internal chains that enforce that each tile exists in the same state.   The right panel shows an explicit crossing tile in the Chimera lattice that allows red and green chains to propagate through one another, encoding non-planarity.}
\label{fig:tileable}
\end{figure}

  Figure \ref{fig:tileable} gives an example of the tileable embedding strategy:  a tileable embedding of the non-planar graph $\mathrm{K}_5$ is presented, along with an illustration of how the crossing tiles work.   In general we will associate multiple tiles to the same vertex in the graph.  One might ask whether this imposes a constraint that if $H_v$ contains $q$ bits (for each vertex) that $\ell$ be sufficient large that $\mathrm{K}_q$ can be embedded in a single tile.  In all of the tileable embeddings we describe below, this will indeed be the case, but it may not be strictly necessary.   Also note that for two vertices $u$ and $v$, a single $H_{uv}$ is added to the tiled QUBO, even if there are multiple possible locations for the coupling.

With the three assumptions above, the embedding problem now becomes far more simple.  We look for minor embeddings of the graph associated with the combinatorial  problem, $G=(V,E)$, into a cubic lattice in two dimensions, subject to the caveat that we may associate some tiles in the square lattice to crossing tiles.    Once we develop the QUBOs that  fit into the fundamental tiles themselves, we can embed \emph{arbitrary} graph combinatorial problems.   The same amount of computational effort is expended by the minor embedding problem in all cases.   Preliminary numerical tests have demonstrated these tileable embeddings for D-Wave's Chimera graph can be found over an order of magnitude faster than conventional embeddings when $\ell \ge 3$.    So in addition to  improving the size of the embeddings for many problems, they can also be found  with far less computational effort.

\subsection{Simultaneous Tiling of Two Problems}\label{sec:2prob}
One technique that will arise in Section \ref{sec:ham}, but which we expect may be of  broad applicability, is a method for tiling two problems (possibly on two separate graphs).   The idea is rather simple, and largely relates to assumption (\emph{iii}) in the previous subsection.   Consider trying to embed a QUBO of the form \begin{equation}
H = H^{(1)}(x_i) + H^{(2)}(y_i) + \sum_{i=1}^N A_i x_i y_i,  \label{eq:2embed}
\end{equation}
where $H^{(1)}$ and $H^{(2)}$ are two constraint type Hamiltonians (e.g. a unary constraint as in Section \ref{sec:tree}), and the $A_i$ are local constraints.       In the problems we will study in this paper, one of the constraints $H^{(1)}$ is generally easy and flexible to efficiently embed, while the other $H^{(2)}$ might be hard.

\begin{figure}[t]
\centering
\includegraphics[width=3in]{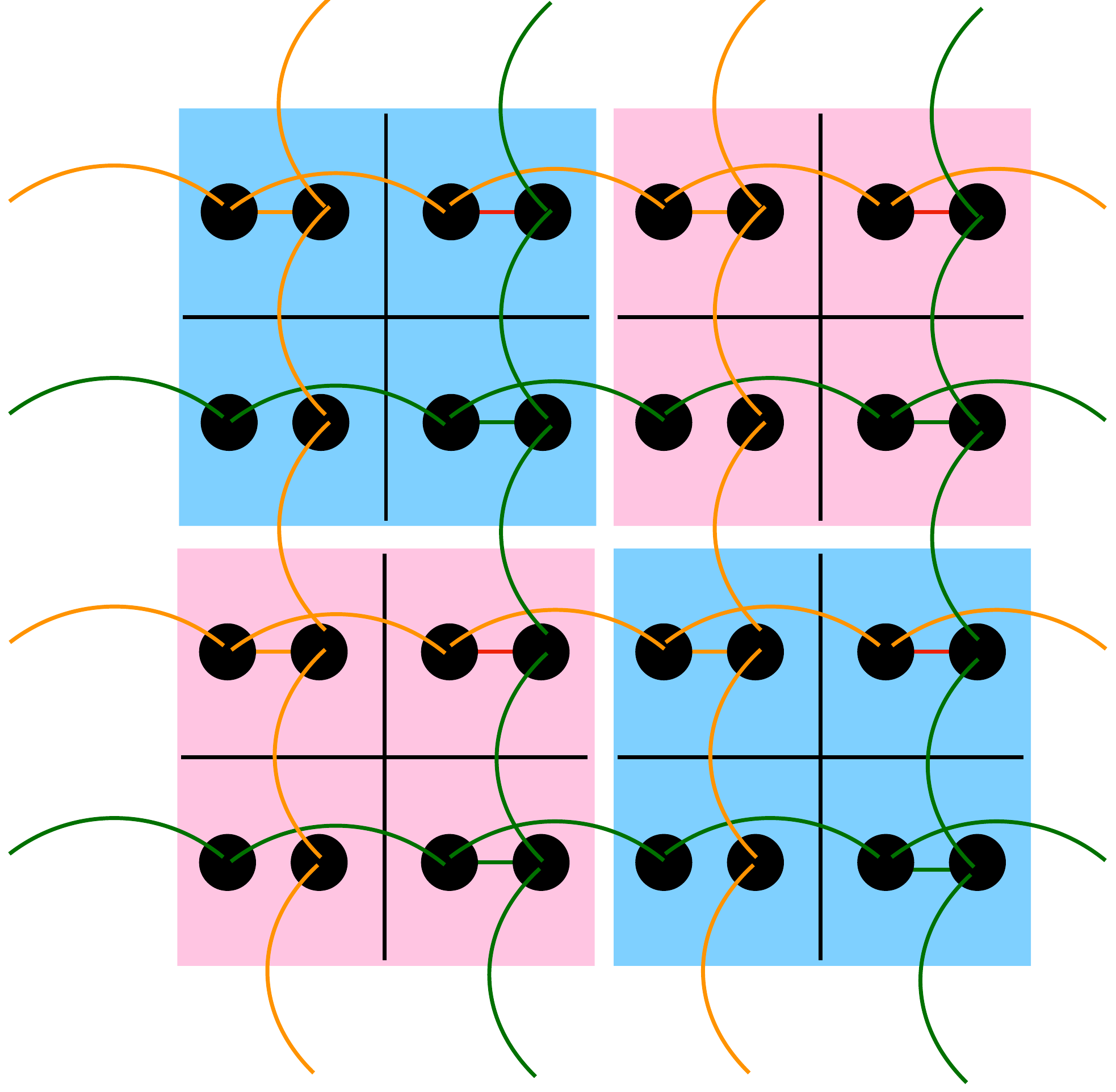}
\caption{A strategy for embedding Hamiltonians of the form (\ref{eq:2embed}).  The blue/pink squares refer to alternating ``supertiles" that encode different vertices $i$.   The red bonds denote possible couplings of  the form $A_i x_i y_i$;  the orange bonds encode couplings in $H^{(1)}$, and the dark green bonds encode couplings in $H^{(2)}$.}
\label{fig:embed2}
\end{figure}

One strategy for embedding such a Hamiltonian is sketched in Figure \ref{fig:embed2}.    Suppose that we find two \emph{separate} embeddings for $H^{(1)}$ and $H^{(2)}$ on an $L\times L$ lattice graph.   Assuming that the embedding for $H^{(1)}$ can be flexibly modified, we also assume that there exists a location $\mathbf{r}_i$ in the $L\times L$ lattice where vertex $i$ sits in the embedding of both $H^{(1)}$ and $H^{(2)}$.   Then we can embed the full Hamiltonian (\ref{eq:2embed}) in a $2L\times 2L$ lattice as shown in Figure \ref{fig:embed2}, by thinking of the $2L\times 2L$ lattice as $L^2$ $2\times 2$ ``supertiles", and placing $H^{(1)}$ in  the upper left and $H^{(2)}$ in the lower right of each supertile.   Coupling $x_i$ to $y_i$ is easily done by adding appropriate couplings in off-diagonal squares  in the supertiles located at appropriate positions  $\mathbf{r}_i$.   

%[IN LATER REVISIONS:  MAKE ANALOGY TO SUPERCELL METHODS FROM 2016-7 LITERATURE?]

\section{Graph Coloring}\label{sec:color}
We now turn to a discussion of the graph coloring problem: given an undirected graph $G=(V,E)$ with $|V|=N$ vertices, can we assign one of $q$ colors to each vertex $v\in V$ such that if $(uv)\in E$,  $u$ and $v$ are assigned different colors?    This problem  is NP-hard for finite $q$.   

A simple Hamiltonian for this problem is \cite{lucas} \begin{equation}
H = \sum_v \left(1-\sum_{i=1}^q x_{v,i} \right)^2  + \sum_{(uv)\in E} \sum_{i=1}^q x_{u,i}x_{v,i}.  \label{eq:oldgraphcoloring}
\end{equation}
The first term encodes a unary constraint that every vertex must have a unique color;  the second term enforces that two adjacent vertices are not assigned the same color.    

Finding a tileable embedding for the graph coloring problem is relatively straightforward.  We simply need to enforce the unary constraint in (\ref{eq:oldgraphcoloring}) within each tile, and subsequently allow for the constraint that neighboring tiles do not have identical colors.   One constraint on tileability is that each tile must consist of $\ell\times \ell$ cells, where \begin{equation}
\ell \ge \left\lceil \frac{q}{\min(e_{\mathrm{h}},e_{\mathrm{v}})}\right\rceil,
\end{equation}
which arises because we will need to compare all $q$ colors between adjacent tiles in the graph.

To say more, we need to study a specific lattice graph.  A natural example is the ($J=4$) Chimera graph used in currently realized experimental devices.  In this case, $\ell = \lceil \frac{q}{4}\rceil$, and beyond the need to implement chains as part of minor embedding, we will not introduce any new ancilla bits.   In fact, when $\frac{q}{4}$ is an integer, the optimal tileable embedding will correspond to assigning vertices within each tile as shown in Figure \ref{fig:K44complete}:  namely each $x_{v,i}$ is propagated through each tile along one vertical and one horizontal chain.   Let $s^i$ and $r^i$ denote spins in the two columns of the Chimera tiles, as shown in Figure \ref{fig:K44}; also define for convenience the following two functions within a single cell of Chimera:    
\begin{subequations}\begin{align}
  H_{\mathrm{diag}}(s^i, r^i) &= \left(\sum_{i=1}^4 s^i\right)\left(\sum_{i=1}^4 r^i\right) - 2\sum_{i=1}^4 s^ir^i + 2\sum_{i=1}^4 \left(s^i+r^i\right), \\
  H_{\mathrm{off}}(s^i, r^i) &= \frac{1}{2}\left(\sum_{i=1}^4 s^i\right)\left(\sum_{i=1}^4 r^i\right) + 2\sum_{i=1}^4 \left(s^i+r^i\right). \\
  \end{align}\end{subequations}  
  When $q\le 4$, a tileable Hamiltonian of the form (\ref{eq:tileableH}), compatible with the hardware constraint (\ref{eq:couplingrange}),  is \begin{subequations}\begin{align}
H_v &= \sum_{\text{tiles}} \frac{1}{2} H_{\mathrm{diag}}  + \cdots, \\
H_{uv}^{\mathrm{hor}} &=  \frac{1}{2}\sum_{i=1}^4 \left(s^i_u+1\right)\left(s^i_v+1\right), \\
H_{uv}^{\mathrm{vert}} &=  \frac{1}{2}\sum_{i=1}^4 \left(r^i_u+1\right)\left(r^i_v+1\right), 
\end{align}\end{subequations}
where $\cdots$ denotes the chain terms to propagate a single vertex through multiple tiles.   The net Hamiltonian has classical energy gap \begin{equation}
\Delta_{q\le 4} = 2  \label{eq:Deltaqle4}
\end{equation}
between a ground state and an excited state.   At the classical level, this gap is optimal for any tileable Hamiltonian for the graph coloring problem: see Appendix \ref{app:chimeracolor}.

Chains that enforce the same values of $s^i$ between tiles corresponding to the same vertex are propagated easily:  if $a$ and $b$ are adjacent horizontal tiles, \begin{equation}
H_{\text{hor. chain}} = -\sum_{i=1}^4 s^i_a s^i_b,  \label{eq:Hhorchain}
\end{equation}
while if they are adjacent vertically:
\begin{equation}
H_{\text{vert. chain}} = -\sum_{i=1}^4 r^i_a r^i_b.  \label{eq:Hvertchain}
\end{equation}
The chain Hamiltonians also have an energy gap of 2, and leave (\ref{eq:Deltaqle4}) unchanged.

When $q>4$, the tileable Hamiltonian that has largest classical energy gap is \begin{subequations}\begin{align}
H_v &= \frac{2}{3} \sum_{\mathrm{tiles}}\left[ \frac{1}{2}\sum_m H_{\mathrm{diag}}\left(s^i_{m,m},r^i_{m,m}\right) + \sum_{m\ne n} H_{\mathrm{off}}\left(\sum_{i=1}^4 s^i_{m,n},\sum_{i=1}^4r^i_{m,n}\right)\right. \notag \\
&\;\;\;\;\;\; \left.- \sum_{i=1}^4\sum_{m=1}^{\ell -1}\sum_{n=1}^\ell s^i_{m,n}s^i_{m+1,n}- \sum_{i=1}^4\sum_{n=1}^{\ell -1}\sum_{m=1}^\ell r^i_{m,n}r^i_{m,n+1}\right] + \cdots,   \label{eq:optHvqg4} \\
H_{uv}^{\mathrm{hor}} &=  \frac{1}{3}\sum_{i=1}^4 \sum_{m=1}^\ell \left(s^i_{u,\ell,m}+1\right)\left(s^i_{v,1,m}+1\right), \\
H_{uv}^{\mathrm{vert}} &=  \frac{1}{3}\sum_{i=1}^4 \left(r^i_{u,m,\ell}+1\right)\left(r^i_{v,m,1}+1\right).
\end{align}\end{subequations}
In the latter two equations we have assumed that tile $u$ is to the top left and tile $v$ is to the bottom right.   The $\cdots$ denote chains between tiles of the same vertex, which are propagated analogously to (\ref{eq:Hhorchain}) and (\ref{eq:Hvertchain}).   The classical energy gap to an excited state is  \begin{equation}
\Delta_{q > 4} = \frac{4}{3},  \label{eq:boundgapqg4}
\end{equation}
which we show is optimal in Appendix \ref{app:chimeracolor}.

It is possible to improve the graph coloring embedding further by observing that the constraints $H_{uv}$ can also be encoded in a crossing tile.\footnote{I thank Kelly Boothby and Aidan Roy for pointing this out.}  This can reduce the number of tiles necessary to encode the embedding but does not enhance the classical energy gap.

\section{Hamiltonian Cycles}\label{sec:ham}
In this section, we address how to construct the tileable embedding for the Hamiltonian cycles problem: does there exist a cycle in an (undirected) graph $G=(V,E)$ that contains every single vertex?
\subsection{Intersecting Cliques}
Before discussing the tileable embedding, however, it is worth reviewing the way that this problem  has been embedded thus far.   The key subtlety in encoding Hamiltonian cycles is the global nature of the constraint:  each vertex must be connected in a single cycle of length $L$.  The current method for addressing this was developed in \cite{lucas}:  let $x_{v,j}$ be a bit that is 1 if vertex $v$ is located in position $j$ in a tentative cycle, and 0 otherwise.   Then the constraint that every vertex is visited exactly  once can be encoded with \begin{equation}
H_{\mathrm{IC}} = \sum_{v=1}^N \left(1-\sum_{j=1}^Nx_{v,j}\right)^2 + \sum_{j=1}^N \left(1-\sum_{v=1}^Nx_{v,j}\right)^2 .\label{eq:ICH}
\end{equation}
The ground states of (\ref{eq:ICH}) encode valid permutations, and we have denoted $|V|=N$.    The constraint that the permutation encodes a Hamiltonian cycle is achieved by writing \begin{equation}
H =  H_{\mathrm{IC}} + \sum_{uv\notin E} \sum_{j=1}^N x_{u,j} x_{v,j+1}.  \label{eq:ICH2}
\end{equation}
Inside bit indices, $j=N+1$ and $j=1$ are identified with one another.   $uv$ and $vu$ are counted as distinct edges in the sum above.

In \cite{rieffel}, the intersecting cliques graph \begin{equation}
\mathrm{IC}_{N,N} = \left(\lbrace ij\rbrace,  \lbrace (ij,kl) \; | \; i=k \text{ or } j=l\rbrace  \right)
\end{equation}
was defined.   $\mathrm{IC}_{N,N}$ is the graph that describes the connectivity of the QUBO (\ref{eq:ICH}), and has a few noteworthy features.  It is rather sparse:  there are $N^2$ vertices, but the degree of every vertex is $2(N-1)$.  On the other hand, the diameter of the graph is 2.    This makes it rather challenging to find good minor embeddings for $\mathrm{IC}_{N,N}$ into a lattice.   In fact, we claim that $L = \mathrm{O}(N^2)$ for any minor embedding of $\mathrm{IC}_{N,N}$ into a lattice.   The key observation is that to embed into a lattice, we must be able to write $V$ as a union of three disjoint sets: \begin{equation}
V = V_+ \cup V_- \cup V_*
\end{equation}
where $V_+$ corresponds to vertices whose chains lie entirely in the top half of the lattice,  $V_-$ corresponds to the bottom half,  and $V_*$ corresponds to vertices whose chains lie in both the top and bottom.   If $|E_{+-}|$ is the number of edges connecting a vertex in $V_+$ to a vertex in $V_-$, then \begin{equation}
|E_{+-}| + |V_*| \le e_{\mathrm{v}}L.
\end{equation}
For $\mathrm{IC}_{N,N}$, we expect that the optimal division of $V$ is as follows (though we have not found a proof).   Let $V_+$ consist of vertices $(vj)$ where $v$ and $j$ are both odd;  let $V_-$ correspond to $v$ and $j$ both even, and $V_*$ correspond to the remainder.   Then $|E_{+-} | = 0$ and $V_* = \frac{1}{2}N^2$.    This suggests that $\mathrm{IC}_{N,N}$ is (up to a constant prefactor) just as hard to minor embed into a lattice as a complete graph with an equivalent number ($N^2$) of  vertices.

\subsection{Tileable Embedding for Intersecting Cliques}\label{sec:perm}
As we have seen previously, in many problems the embedding can be greatly improved by looking for treelike embeddings of constraints.     This can also be accomplished for (\ref{eq:ICH}).   We will encode the first set of constraints in (\ref{eq:ICH}) -- fixed $i$ and variable $j$ -- locally.   This generally requires a complete graph.   The  second set of constraints -- fixed $j$ and variable $i$ -- will \emph{simultaneously} be embedded as a global treelike constraint.     We must thread $N$ trees through the lattice simultaneously.   This can be accomplished using logic analogous to Section \ref{sec:2prob}, but generalized to $N$ problems.   Alternatively, in the same way in which we encoded two simultaneous branches of the unary constraint as for the tree in Section \ref{sec:tree}, we must now embed $N$ constraints.  

\begin{figure}[t]
\centering
\includegraphics[width=4in]{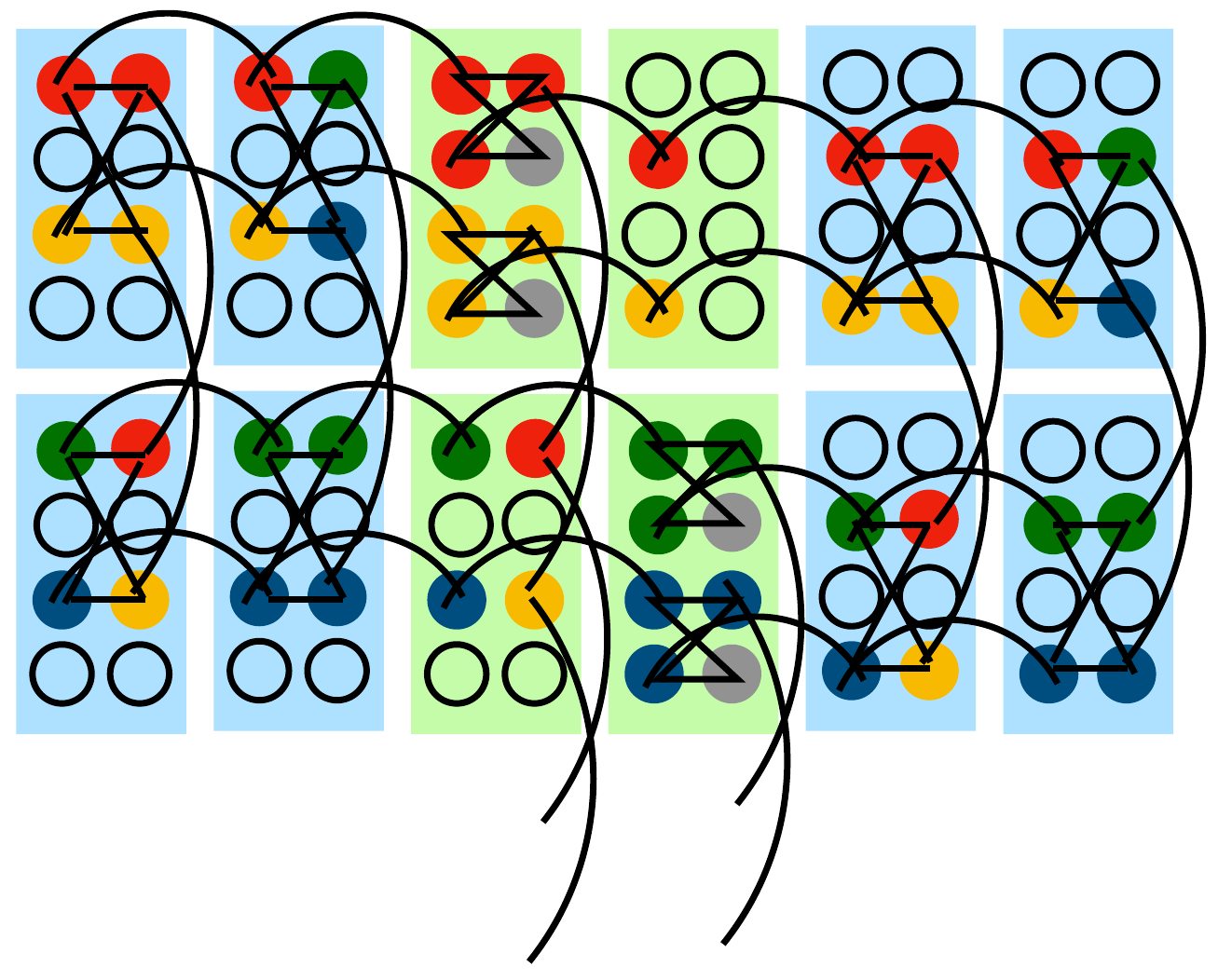}
\caption{Part of the treelike embedding of a QUBO whose ground states are  permutations on $N=4$ objects.   The 4  colors of vertices denote the 4 possible values of $j$;  the two blue tiles encode the constraints at fixed $i$, while the green tile helps to encode the treelike global constraints.}
\label{fig:treeIC}
\end{figure}

This is achievable, but it is helpful to see how this embedding works with a specific example.  In Figure \ref{fig:treeIC} we show the topmost $6\times 2$ sublattice of a $6\times 6$ Chimera lattice that encodes the permutation constraint on $N=4$ objects.     Observe that now each tile of the embedding is larger than a single unit cell of the lattice:  in this case, we use $2\times 2$ tiles.   More generally, we can encode a permutation constraint using tiles of length $\frac{N}{2}$.   This number is bounded by our ability to merge two constraints, which requires  one $\mathrm{K}_{2,2}$ sublattice for each $j$, along with our ability to  propagate $N$ chains along a single side of a tile.

Let us now compare the size of the treelike embedding to the size of a conventional embedding.  The size of the conventional embedding is going to be $L\approx \frac{1}{4}N^2$ since we just  encode a complete graph on $N^2$ bits.   In contrast, the treelike embedding requires \begin{equation}
L \approx   \frac{N}{2}\left(2\sqrt{N} -  1\right).
\end{equation}
If we want to embed a permutation  constraint with $N=16$, the complete embedding requires $L=64$ while the treelike embedding requires $L=56$.   This value of $L$ is larger than what is currently accessible in hardware  but may be achievable in the near future.    A fully optimized permutation constraint might be competitive for smaller values of $N$ and $L$.

\subsection{Embedding Hamiltonian Cycles}\label{sec:hamcons}
To encode the Hamiltonian cycles problem, we will take a different approach than in (\ref{eq:ICH2}).   Let us consider the following naive Hamiltonian:  \begin{align}
H &= \sum_{v=1}^N \left[ \left(1-\sum_{j=1}^N x_{v,j}\right)^2 +  \left(1-\sum_{u:(uv)\in E}  z_{v,uv}\right)^2 + \sum_{u:(uv)\in E} \left(z_{v,uv} - \sum_{j=1}^N z_{v,uv,j}\right)^2 \right] \notag \\
& \;\;\;\;\;\; + \sum_{v=1}^N  \sum_{u:(uv)\in E} \sum_{j=1}^N \overline{\mathbb{I}}\left(z_{v,uv,j} = x_{v,j}x_{u,j+1}\right).  \label{eq:83ham}
\end{align}
States on which $H=0$ encode solutions to the Hamiltonian cycles problem on the graph because the only way that a $z_{v,uv}=1$ is if one of the $z_{v,uv,j}=1$, which occurs when there is a neighbor $u$ of each vertex $v$ for which the color index $j$ (i.e., position in the cycle) has increased by 1 (mod $N$).   If two vertices took the same value of $j$, then since there are $N$ colors there would be a vertex that could not have a neighbor whose $j$ was 1 larger.   This would imply an energy penalty, either because all $z_{v,uv}=0$ or because one of the constraints on $z_{v,uv}$ or $z_{v,uv,j}$ would be violated.    Thus we conclude that states with $H=0$ assign each vertex a unique value of $j$ and that vertices $j$ and $j+1$ are neighbors, which is indeed sufficient for the Hamiltonian cycles problem.

In order to encode this $H$ on a lattice graph, we observe that the only terms in $H$ which connect bits associated with vertex $v$ to vertex $u$ are found in the second line of (\ref{eq:83ham}).   Such terms only exist for pairs $(uv)\in E$.   Thus if  we find a tileable embedding of the graph $G=(V,E)$, and can also construct suitable tiles, we can embed (\ref{eq:83ham}).

To construct these tiles, our primary goal is now to further improve upon the second constraint in the first line of (\ref{eq:83ham}).   The reason  is as follows:   suppose that each tile in the lattice was associated to a unique vertex -- namely, the graph $G$ is itself a square lattice.   Then the number of bits in every tile is $\mathrm{O}(N)$ (we will soon precisely specify how many), and we have $\mathrm{O}(N)$ tiles.    Assuming the worst case scenario -- each tile encodes a complete graph (as in the coloring problem) -- we would then find that the cycles problem could be encoded on a graph of size $L\sim NL_G \sim N^{3/2}$, which parametrically improves upon the scaling $L\sim N^2$.  

\begin{figure}[t]
\centering
\includegraphics[width=4in]{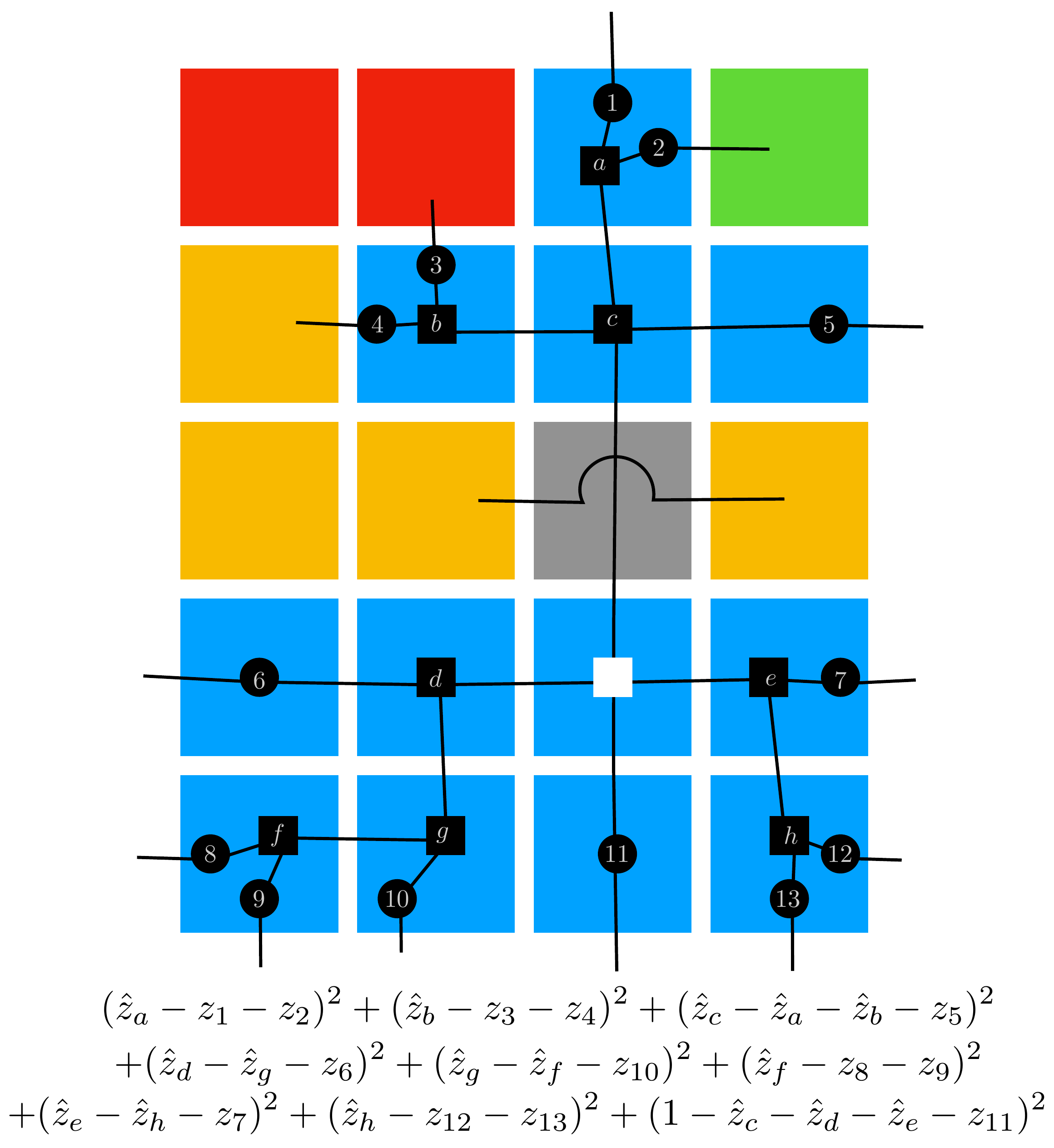}
\caption{An internal treelike structure can be used to efficiently embed a unary constraint in an arbitrary collection of tiles corresponding to a single vertex $v$ (in this case, shown in blue).   Circular nodes correspond to $z_{v,uv}$, and square black nodes correspond to internal auxiliary bits $\hat{z}_{v,a}$ which encode the overall unary constraint.  The white node denotes the ``root" of the tree encoding the unary constraint.  A specific Hamiltonian (where we have replaced all bits on chains with a single effective bit) is written below the diagram.  As in Figure \ref{fig:tileable}, the gray tile represents a crossing tile to encode non-planarity in the graph $G$.}
\label{fig:tree8}
\end{figure}

Our goal is to embed the unary constraint on $z_{v,uv}$ such that we do not propagate chains of the $\mathrm{O}(N)$ $z_{v,uv,j}$ bits throughout every $v$ tile in a generic tileable embedding.    We can achieve this as shown in Figure \ref{fig:tree8}, using methods analogous to Section \ref{sec:tree}.  In each tile, we add a new bit that enforces a unary constraint among the $z_{v,uv}$ bits within that tile.    The overall unary constraint on $\sum z_{v,uv}$ is imposed by treelike couplings between the auxiliary degrees of freedom.  Note that we need  add  only one $z_{v,uv}$ for a given edge $uv$, even if there are multiple $u$ and $v$ tiles that are adjacent in the embedding.    A sloppy embedding of (\ref{eq:83ham}) is possible, improving only the unary constraint on $\sum z_{v,uv}$, by adding 5 bits per tile.

There is one final issue that we need to address.  The second line of (\ref{eq:83ham}) involves an indicator function which cannot be directly embedded as it involves a coupling between three bits directly.   This can be solved by the following trick:
\begin{equation}
\overline{\mathbb{I}}\left(z_{v,uv,j} = x_{v,j}x_{u,j+1}\right) =  \frac{1}{2}\left[4z_{v,uv,j}+ 2x_{v,j}x_{u,j+1} - 3z_{v,uv}(x_{v,j}+x_{u,j+1})\right].  \label{eq:3bitcons}
\end{equation}
This has not introduced any auxiliary bits.   Because $x_{u,j+1}$ now couples to two bits within a given $v$ tile, we must have an ancilla bit inside of the $u$ tile to represent $x_{u,j+1}$.    

Combining (\ref{eq:83ham}),  (\ref{eq:3bitcons}) and the algorithm of Figure \ref{fig:tree8}, we have provided an explicit algorithm for embedding the Hamiltonian cycles problem.    For ease of presentation, let us now explicitly count the lattice size required to embed this problem on a Chimera lattice.   As we saw in Section \ref{sec:color}, we can embed $\mathrm{K}_{4\ell}$ in an $\ell\times\ell$ tile.    If $L_G$ is the number of tiles in each dimension required in order to embed the graph $G$, we first estimate that $L \le \frac{L_G}{4} 9(N+1)$.   
$9N+9$ is the number of bits required per tile:  $N$ are $x_{v,j}$;  $4(N+1)$ count the possible $z_{v,uv}$ and $z_{v,uv,j}$;  $4N$ count the possible $x_{u,j+1}$, and 5 count the auxiliary bits for the internal treelike unary constraint.    In fact, there is an immediate improvement.   Consider \emph{doubling} the number of tiles in each (of two) dimensions, but forcing a maximum of one $z_{v,uv}$ per tile.    This doubles $L$, but reduces the number of tiles to $3N+5$ (now $z_{v,uv}$ can simply be taken to be one of the 5 bits encoding the local unary constraint on all $z_{v,uv}$).   Thus we find an improvement to
\begin{equation}
L \le \frac{L_G}{2} (3N+5).  \label{eq:Lham}
\end{equation}
As we can embed some graphs with $L = \mathrm{O}(\sqrt{N})$, we conclude that there exist graphs for which the Hamiltonian cycles problem can be embedded using $L=\mathrm{O}(N^{3/2})$, which is parametrically better in the large $N$ limit.

Unfortunately, this particular embedding is presently more useful as a theoretical construction than a practical algorithm.   Using the Hamiltonian (\ref{eq:ICH2}) and embedding via a complete graph, it is possible to embed the Hamiltonian cycles problem in Chimera using $L=\frac{1}{4}N^2$.    We find that $N\sim 45$ is the smallest $N$ for which (\ref{eq:Lham}) represents an improvement (assuming $L_G=7$).  This corresponds to $L=490$ for the tileable embedding, versus $L=504$ for the complete embedding.   Both values require an order of magnitude larger lattice than will be achievable in the near term future.

%On the Chimera graph, the value of $L$ may be reduced analogously to the discussion in Section \ref{sec:tree}.   We need to (\emph{i}) embed $\mathrm{O}(L_G^2)$ cells and (\emph{ii}) propagate $N$ unary constraints between each of them.    In order, these requirements can be embedded in $L_G \times \frac{5}{4}(N+1)$ rows of Chimera and $L_G\times N$ rows, respectively.   Thus $L \approx \frac{9}{4}NL_G \gtrsim \frac{9}{4}N^{3/2}$.    Using the embedding of a complete graph, it is possible to embed the Hamiltonian cycles problem using $L=\frac{N^2}{4}$ on the Chimera graph.   Thus the new embedding of Hamiltonian cycles is more effective when $N\ge 81$, on a graph which can be optimally embedded.  This corresponds to $L\sim 1600$, which is far from being realized using current technologies.  Since graphs which can be optimally embedded are planar they are likely not the most interesting graphs to study the cycles problem on, and so this scaling may be worse on more non-trivial graphs.   Despite this unfortunate outlook for the short term, our results  demonstrate novel asymptotic scaling and it may be possible to find more efficient embeddings beyond what we have presented. 

\section{Conclusion}\label{sec:conc}
The primary goal of this work has been to demonstrate a number of novel embedding strategies that lead to parametrically enhanced scalings in the size of combinatorial QUBOs embeddable within a given lattice.   Some of the strategies described above are asymptotically optimal, at least as measured by the problem size that can be embedded on a given graph size.   It  will be interesting to further improve upon the methods developed here, reducing embedding sizes further by filling in unused portions of the graph (as briefly discussed in Section \ref{sec:treeopt}).  We leave such constant size improvements to future work.

Our results have direct implications on the combinatorial problems for which quantum annealers may be most promising.  In particular, we have found that in addition to carefully tailored (sub)problems that manifestly embed onto the Chimera graph (e.g.,  Ising spin glass with Chimera topology \cite{boixo14}),  NP-hard optimization problems including knapsack and partitioning problems also embed rather efficiently due to the simple global nature of the constraint.    This result is general and relies only on spatial locality of the lattice on sufficiently large scales.   A more detailed study on quantum annealing of such problems is worthwhile.

Embeddings that require long chains tend to perform worse on D-Wave hardware \cite{bian}.   Unfortunately, due to the constraints from spatial locality, this problem is in some respects unavoidable.   For example, in the treelike embeddings described above, we have only addressed this problem partially.  Indeed, there are far fewer chains representing individual ancilla bits;  however, there are many new auxiliary bits.   There can continue to exist widely separated physical bits in the lattice whose logical counterparts interact in the abstract QUBO.   It would be interesting to understand whether our new approach has in any way ameliorated the problem of spatial locality and chain propagation in highly connected QUBOs such as partitioning and knapsack.

\addcontentsline{toc}{section}{Acknowledgements}
\section*{Acknowledgements}
I thank Kelly Boothby, Fiona Harrington, Catherine McGeoch, and Aidan Roy for helpful feedback.    This work was supported by D-Wave Systems, Inc.

\begin{appendix}
\section{A Cartoon of Exponentially Small Spectral Gaps}\label{app:gap}
Consider a classical combinatorial problem to find  the minimum of a function $H(x_1,x_2,\ldots, x_N)$ of binary variables $x_i$.  Without loss of generality, we define the $x_i$ such that a ground state of $H$ is $x_i=0$.   For simplicity, we will  suppose that this ground state is unique.

 A typical QA protocol is as follows:  define a time-dependent Hamiltonian \begin{equation}
H(t) = \left(1-\frac{t}{\tau_{\mathrm{QA}}}\right) H_{\mathrm{drive}} + \frac{t}{\tau_{\mathrm{QA}}} H(Z_1,Z_2,\ldots, Z_N),	
\end{equation}
where \begin{equation}
H_{\mathrm{drive}} \propto -\sum_{i=1}^N X_i 
\end{equation}and \begin{subequations}\begin{align}
X_i |x_1 \cdots 0_i \cdots x_N \rangle &= |x_1 \cdots 1_i \cdots x_N \rangle, \\
X_i |x_1 \cdots 1_i \cdots x_N \rangle &= |x_1 \cdots 0_i \cdots x_N \rangle, \\
Z_i |x_1 \cdots 0_i \cdots x_N \rangle &= 0, \\
Z_i |x_1 \cdots 1_i \cdots x_N \rangle &= |x_1 \cdots 1_i \cdots x_N \rangle
\end{align}\end{subequations}
denote quantum operators acting on a tensor product Hilbert space of $N$ two-state quantum systems.  We assume that we  can prepare the quantum state $|\Psi\rangle$ in the ground state of $H_{\mathrm{drive}}$ for $t<0$; we will denote this ground state with\begin{equation}
|\Omega\rangle = \frac{1}{2^{N/2}} \sum_{x_1,\ldots, x_N \in \lbrace 0,1\rbrace} |x_1\cdots x_N\rangle.
\end{equation}

Let us modify this protocol somewhat.  Let us define \cite{alan1407} \begin{equation}
H_{\mathrm{drive}} = 1-|\Omega\rangle\langle \Omega|,  \label{eq:Hdriveapp}
\end{equation}
and let us suppose that \begin{equation}
H(x_1,\ldots,x_N)  = 1  - \prod_{i=1}^N(1-x_i).
\end{equation}
The quantum state evolves in a two-dimensional subspace of the many-body Hilbert space spanned by \begin{subequations}\begin{align}
|\downarrow \rangle &= |0_1\cdots 0_N\rangle, \\
|\uparrow \rangle &=  \frac{1}{\sqrt{2^N-1}}\sum_{(x_1,\ldots, x_N) \ne (0,\ldots,0)} |x_i\rangle.
\end{align}\end{subequations}
 Define $\epsilon = 2^{-N}$.   At time $t=0$, the quantum state is \begin{equation}
 |\Psi(0)\rangle = |\Omega\rangle = \sqrt{\epsilon} |\downarrow\rangle + \sqrt{1-\epsilon}|\uparrow\rangle,
 \end{equation}
 and the Hamiltonian is given by \begin{equation}
 H = -(1-s)\epsilon|\downarrow\rangle\langle \downarrow | - (1-s)\sqrt{\epsilon(1-\epsilon)} (|\downarrow \rangle\langle \uparrow | + |\uparrow \rangle\langle \downarrow |)  + (s - (1-s)(1-\epsilon))|\uparrow\rangle\langle \uparrow |
 \end{equation}
 where $s=t/\tau_{\mathrm{QA}}$.   The minimal spectral gap of $H$ is \begin{equation}
 \Delta = \sqrt{\epsilon}
 \end{equation}
 and occurs when $s=1/2$.   Using the Landau-Zener formula, we expect that \begin{equation}
 \tau_{\mathrm{QA}} = \frac{1}{\epsilon}.  \label{eq:tauQAapp}
 \end{equation}
 Note that this formula is completely insensitive to the hardness of the problem.   Using an optimal annealing schedule, \cite{alan1407} could improve this time to \begin{equation}
  \tau_{\mathrm{QA}} = \frac{1}{\sqrt{\epsilon}}.   \label{eq:tauQAapp2}
 \end{equation}
 The scaling (\ref{eq:tauQAapp2}) is the quadratic speedup noted in the introduction.
 
In \cite{alan1407}, it was shown that (\ref{eq:tauQAapp2}) is actually necessary in the quantum annealing of the random energy model \cite{derrida} with a simple $H_{\mathrm{drive}}$.  The energy spectrum of the random energy model is actually equivalent to the energy spectrum of a rather boring Hamiltonian such as (\ref{eq:Hdriveapp}).   Without a carefully chosen QA algorithm, we expect poor performance of QA.   Part of choosing a good QA algorithm, using available technology, involves finding a clever minor embedding, which is the focus of the main text of this paper.

\section{Tileable Embeddings for Graph Coloring on Chimera Lattices}\label{app:chimeracolor}
  In this appendix we find upper bounds on that the classical energy gaps for tileable embeddings of graph coloring, and ultimately show that the Hamiltonians presented in Section \ref{sec:color} have optimal classical gap. 
  
  \subsection{$q\le 4$}  
When $q\le 4$, we can encode $H_v$ in a single tile of the Chimera.  Let $a$ and $b$ be adjacent tiles.  We can restrict the form of $H_v$ and $H_{uv}$ using the permutation symmetry $\mathrm{S}_q$.  We look for $H$ of the form \begin{subequations}\label{eq:qle4H}\begin{align}
  (H_v)_a &= \lambda \left[A \left(\sum_{i=1}^4 s^i_a \right)\left(\sum_{i=1}^4 r^i_a \right) + B\sum_{i=1}^4 s^i_a r^i_a + C \sum_{i=1}^4 (s^i_a+r^i_a)\right], \\
    (H_{uv})_{ab} &= D \sum_i (s^i_a+1) (s^i_b+1).
  \end{align}\end{subequations}
  Without loss of generality in the second line, we have assumed that the $x$ bits are coupled between two adjacent horizontal vertices;  if the bits are instead adjacent vertically, then the $y$ bits are coupled.   From (\ref{eq:couplingrange}), \begin{subequations}\label{eq:qle4cons}\begin{align}
|A|, |A+B|, |D| &\le 1, \\
|C| &\le 2.
\end{align}\end{subequations}
Note that choosing the Hamiltonian to take the form (\ref{eq:qle4H}) guarantees that the ground states correctly solve the graph coloring problem.  

In the absence of other vertices (i.e., $\lambda=1$), $H_v$ can be chosen to saturate all bounds in (\ref{eq:qle4cons}):  \begin{subequations}\begin{align}
A &= 1, \\
B &= -2, \\
C &= 2.
\end{align}\end{subequations} 
It is simple to check that the spectral gap of $H_v$ is \begin{equation}
\Delta_{q\le 4}^{H_v} = 4.  \label{eq:deltaq4Hv}
\end{equation}  In the presence of other tiles, there may be two neighboring vertices for which $H_{uv}$ must be imposed.   This implies that \begin{equation}
\lambda C+2D = 2(D+\lambda) = 2
\end{equation}
on the optimal mapping.    To determine the optimal value of $\lambda$ (and thus $D$), we observe the following.  If $D\rightarrow 0$, then the spectral gap of $H$ vanishes because we can color two adjacent vertices the same color with vanishing penalty.  If $\lambda \rightarrow 0$, then the penalty for assigning each vertex a unique color vanishes.  Thus the optimal spectral gap will occur when $0<\lambda=1-D<1$.  To find the worst case scenario, we consider the following:  if we fail to color one tile properly, we will find an energy penalty of $4\lambda$ (using (\ref{eq:deltaq4Hv})).   We may also excite the lowest lying excited state of a single $H_{uv}$, which has an energy gap of 4.   Thus
\begin{equation}
\Delta_{q\le 4} = \max (4\lambda, 4D) = 2.
\end{equation}
The optimal solution has $\lambda=1/2$.   This directly implies (\ref{eq:Deltaqle4}).

  \subsection{$q>4$}
  We now turn to the case of $q>4$.   We will first find the optimal $H_v$, and then follow our previous logic to fix the intertile couplings and find $\Delta$.
  
  Let us denote $s^i_{m,n}$ and $r^i_{m,n}$ as spins in element $(m,n)$ of the $\ell\times \ell$ Chimera tile.   A simple ansatz for $H_v$ is \begin{align} \label{eq:appB2H}
  H_v &=\lambda \sum_{m=1}^\ell \left[ A \left(\sum_{i=1}^4 s^i_{m,m}\right)\left(\sum_{i=1}^4 r^i_{m,m}\right) +  B\sum_{i=1}^4 s^i_{m,m}r^i_{m,m} + C\sum_{i=1}^4 \left(s^i_{m,m}+r^i_{m,m}\right)  \right] \notag \\
  &\;\;\;\;  + \lambda \sum_{m\ne n} \left[ D\left(\sum_{i=1}^4 s^i_{m,n}\right)\left(\sum_{i=1}^4 r^i_{m,n}\right)  +E\sum_{i=1}^4 \left(s^i_{m,n}+r^i_{m,n}\right) \right] \notag \\
  &\;\;\;\; + \lambda F \sum_{m,n=1}^{\ell-1}  \sum_{i=1}^4\left(s^i_{m,n}s^i_{m+1,n} + r^i_{m,n}r^i_{m,n+1} \right).
  \end{align}
  While this is not the most general possible form allowed by the symmetries, the argument that follows below does not depend on the constants $C$, $D$, $E$ etc., being independent of $m$ and $n$.   For simplicity we use the ansatz above to fix simpler notation.
  
  \begin{figure}
  \centering
  \includegraphics[width=1in]{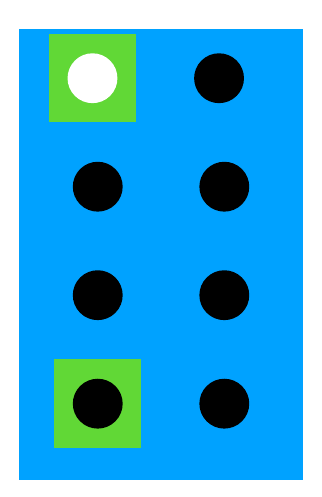}
  \caption{The top right most cell  in  a tile used to encode $H_v$.  White circles denote $+1$ spins in a ground state solution; black denotes $-1$ spins.  The green highlights denote two possible spins which can be flipped to find excited states of the classical Hamiltonian.}
  \label{fig:appB2fig}
  \end{figure}
  
  Rather than directly looking for the couplings $A,\ldots,F$, we are first going to try to understand what the optimal bound on $\Delta^{H_v}_{q>4}$, the spectral gap of $H_v$, might be  on general grounds.   As shown in Figure \ref{fig:appB2fig}, we consider fixing one of the $s^i$ spins in the top most row to be 1;  thus, on the optimal solution, all other spins should be $-1$.   Now consider flipping single spins in the top right most cell of the tile.   Whatever spin we flip we must find an energy penalty of at least $\Delta$.  Two natural choices of spins to flip are highlighted in Figure \ref{fig:appB2fig};  we conclude that \begin{equation}
  \Delta = \min \left(8D - 2E - 2F, 2E - 8D - 2F\right).
  \end{equation}
  Combining these two equations we find \begin{equation}
  2\Delta \le -4F
  \end{equation}
  and since $|F| \le 1$:  \begin{equation}
  \Delta_{q>4}^{H_v} \le 2.  \label{eq:Deltaqg4}
  \end{equation}
  
  Luckily, we will now show that -- up to the overall rescaling factor of $\frac{2}{3}$, which will be explained at the end -- the Hamiltonian $H_v$ given in (\ref{eq:optHvqg4}) saturates this bound.   First, observe that the second line of (\ref{eq:optHvqg4}) above simply enforces that all spins on intercell chains within a tile take the same value.  Secondly, note that we may write \begin{equation}
    H_{\mathrm{off}}(S,R) = \frac{(S+4)(R+4)}{2} - 8 .
  \end{equation}
Finally, $H_{\mathrm{diag}}$ is simply the optimal encoding of a unary constraint in a single tile, and as described above, it has an energy gap of 4 and has ground state where one pair of $(s^i,r^i)$ is $+1$.    $H_{\mathrm{off}}$ is a simple Hamiltonian with an energy gap of 2 that penalizes any state in there is a $+1$ spin in each half of the tile.    
  
  To show that up to rescaling, $H_v$ in (\ref{eq:optHvqg4}) obeys (\ref{eq:Deltaqg4}), we need to evaluate all possible sources of error.  Firstly, note that if any chain is broken, we obtain an energy penalty of 2, from the second line of (\ref{eq:optHvqg4}).   Thus we may assume that all variables in a chain take the same value.   Secondly, observe that $\frac{1}{2}H_{\mathrm{diag}}$ has a gap of 2, and therefore we may also assume that the $s$ and $r$ chains corresponding to the same physical bit take an identical value.   Lastly, let us denote with $N_m \in \lbrace 0, 1\rbrace$  the number of pairs of $+1$ spins in diagonal cell $(m,m)$.   Observe that in the space of states within an energy 2 of the ground states:
  \begin{equation}
  H_v = -2\sum_{m=1}^\ell N_m  + 2\sum_{m\ne n} N_m N_n  + C = 2\left(\sum_{m=1}^\ell N_m\right)^2  - 4\sum_{m=1}^\ell N_m +C ,\;\;\;\; (N_m\in \lbrace 0,1\rbrace)
  \end{equation}
  where $C$ is an unimportant constant offset.   The above $H_v$ takes a minimal value of $C-2$ when exactly one of the $N_m$ is 1, and has an energy gap of 2 (to either $\sum N_m = 0 $ or $\sum N_m = 2 $).   
  Thus, we have found an explicit Hamiltonian $H_v$ that obeys (\ref{eq:Deltaqg4}).     We have additionally checked that this $H_v$ is optimal for $q=8$ and $q=12$ using numerical satisfiability modulo theory solvers \cite{tinelli}.
  
  The last step is to include $H_{uv}$.  Just as before, \begin{equation}
  (H_{uv})_{ab} = G\sum_{n=1}^\ell \sum_{i=1}^4 (s^i_{\ell,n,a}+1)(s^i_{1,n,b}+1)
  \end{equation}
  assuming that the neighboring tiles are horizontally connected.   Rescaling $H_v$ by $\lambda$ as before, we find that \begin{equation}
  \Delta_{q>4} = \min (2\lambda, 4G)
  \end{equation}
  Since the single-site field on an off-diagonal tile cannot have coupling larger than 2: \begin{equation}
  2\lambda + G = 2,
  \end{equation}
  we conclude that $G=4/3$ and $\lambda=2/3$:  thus, we obtain (\ref{eq:boundgapqg4}).

\end{appendix}

\bibliographystyle{unsrt}
\addcontentsline{toc}{section}{References}
\bibliography{isingbib2}

\end{document}